\documentclass[aps,prd,preprintnumbers,groupedaddress,nofootinbib,amssymb,notitlepage,eqsecnum]{revtex4-1}
\usepackage{here}
\usepackage{graphicx}
\usepackage{amsmath}
\usepackage{enumitem}  
\usepackage{bm}
\usepackage{color}
\usepackage[dvipsnames]{xcolor}
\usepackage[utf8]{inputenc}
\usepackage{amsfonts}
\definecolor{refs}{RGB}{245,156,74}
\usepackage[colorlinks=true,hyperfootnotes=true,citecolor=cyan]{hyperref}
\usepackage{comment}
\usepackage{amsmath}
\usepackage{tikz}
\usetikzlibrary{decorations.pathreplacing}
\usepackage{minted}
\usepackage{placeins}

\newcommand{\be}{\begin{equation}}
\newcommand{\ee}{\end{equation}}
\newcommand{\bea}{\begin{eqnarray}}
\newcommand{\eea}{\end{eqnarray}}

\newcommand{\jcap}{JCAP}

\newcommand{\mnras}{Monthly Notices of the Royal Astron. Soc.}

\newcommand{\Dtwo}{D_2}
\newcommand{\Dq}{D_q}
\newcommand{\LCDM}{\Lambda\mathrm{CDM}}

\allowdisplaybreaks

\begin{document}

\title{The Cosmological Correlation Dimension Beyond the Linear Regime: Analytical Approximation and Parameter Sensitivity}

\author{Miguel Aparicio Resco$^{1}$\footnote{{\tt miguel.aparicio@upm.es}}, 
Florencia A. Teppa Pannia$^{2,3,1}$\footnote{{\tt 
f.a.teppa.pannia@usal.es}}
}

\affiliation{
$^{1}$ {\it Departamento de Matem{\'a}tica Aplicada a la Ingenier{\'i}a Industrial, Universidad  Polit{\'e}cnica de Madrid, E-28006 Madrid, Spain.} \\
$^{2}$ {\it Departamento de Estadística, Universidad de Salamanca, E-37008 Salamanca, Spain.} \\
$^{3}$ {\it Instituto Universitario de F\'isica Fundamental y Matem\'aticas~(IUFFyM), Universidad de Salamanca, E-37008 Salamanca, Spain.}
}

\begin{abstract}
The correlation dimension ($D_2$) provides a scale-dependent characterization of the transition towards homogeneity and a complementary perspective on the clustering of large-scale structure. In this work, we study its full dependence on scales and redshift connecting $D_2$ directly to the matter power spectrum of a given cosmological model, and   derive an approximate analytical relation based on an effective cut-off representation of the filtering kernel. We further perform two complementary local (derivative-based) and global (variance-based) sensitivity analyses within the $w_0w_a$CDM model. In general, nonlinear evolution lowers $D_2$  relative to the linear prediction at small scales, with the effect becoming more pronounced at low redshift. Our cut-off approximation, with $k_{\rm cut}=\alpha/r$ and $\alpha\simeq 2.39$, reproduces the exact $D_2$ to better than 1\% for the representative cases considered. The sensitivity analysis identifies $\Omega_m$ and $A_s$ as the dominant parameters for the linear correlation dimension and its nonlinear-to-linear ratio, respectively, while $w_0$ and $w_a$ induce scale- and redshift-dependent changes in the nonlinear contribution. These results provide a framework for characterizing the cosmological scale- and redshift-dependent features of $D_2$ and its nonlinear correction.
\end{abstract}

\date{\today}
\maketitle

\section{Introduction}

The large-scale distribution of matter in the Universe shows a complex network of clusters, filaments, sheets, and voids, usually referred to as the cosmic web. This structure is generated by the gravitational amplification of primordial density fluctuations and is therefore highly inhomogeneous on small and intermediate scales. On sufficiently large scales, however, the matter distribution is expected to approach statistical homogeneity, as required by the cosmological principle. Fractal and multifractal descriptors provide a natural language to quantify this transition, since they measure how the amount of matter, or the number of tracers, enclosed within spheres changes with the scale of observation \cite{Hentschel:1983zhc,Borgani:1994uy,Gaite:2018kbu}.

Among the hierarchy of generalized dimensions, the correlation dimension $\Dtwo(r)$ plays a special role. It is directly related to the scaling of the counts-in-spheres and, equivalently, to the volume integral of the two-point correlation function. For a statistically homogeneous distribution embedded in three spatial dimensions, the average count scales as $N(<r)\propto r^3$ and, therefore, $\Dtwo(r)\to 3$. The scale-dependent clustering of the tracer or matter field is encoded in how $D_2(r)$ approaches this value. In this sense, $\Dtwo(r)$ is not simply a binary test of homogeneity, but a compressed geometric observable that tracks how the cosmic web approaches the homogeneous regime as a function of scale \cite{Yadav:2010cc,Einasto:2020jrp}. 

The correlation dimension has been extensively used as an observational test of large-scale homogeneity. Early and modern analyzes of galaxy and quasar catalogs have generally found a transition from clustered to nearly homogeneous behavior on scales of order $\mathcal{O}(100)\,\mathrm{Mpc}/h$, although the precise value depends on the tracer, survey geometry, estimator, and adopted criterion \cite{Hogg:2004vw,Sarkar_2009,Yadav:2010cc,Scrimgeour:2012wt,Goyal:2024ctd}. These studies have established $\Dtwo(r)$ and the broader spectrum $\Dq(r)$ as robust statistical probes of the cosmic web. Nevertheless, their main emphasis has usually been the empirical determination of the homogeneity scale, rather than the detailed theoretical response of $\Dtwo(r)$ to non-linear structure formation and to changes in the underlying cosmological model.

In this work, we shift the focus from the measurement of a single homogeneity scale to the modeling of the full scale and redshift dependence of the correlation dimension, $\Dtwo(r,z)$. We compute the correlation dimension from the matter power spectrum and study how non-linear evolution modifies it with respect to the linear prediction. This makes it possible to identify the scales and redshifts where non-linear corrections are relevant, to quantify their impact through ratios such as $\Dtwo^{\rm nl}/\Dtwo^{\rm lin}$, and to analyze the sensitivity of $\Dtwo$ to cosmological parameters. We focus quantitatively on parameterization $(w_0,w_a)$ of dynamical dark energy \cite{Chevallier:2000qy,Linder:2002et}, which provides a controlled two-parameter extension of the standard $\LCDM$ model. This choice is also motivated by recent DESI DR2 results, which have shown a preference for a dynamical dark energy equation of state over a cosmological constant when combined with CMB data ($3.1\sigma$) or CMB+SNe data ($2.8-4.2\sigma$) \cite{DESIdr2}. 

In order to quantify the impact of non-linear evolution, we compute the non-linear matter power spectrum using two complementary prescriptions: HMCODE, an augmented halo-model approach calibrated against numerical simulations \cite{Mead:2015yca,Mead:2020vgs}, and Euclid Emulator 2 (EE2), a simulation-based emulator designed to predict the non-linear correction to the matter power spectrum in an extended $w_0w_a$CDM parameter space including massive neutrinos \cite{Euclid:2018mlb,Euclid:2020rfv}. The use of both approaches allows us to compare a physically motivated semi-analytical description of non-linear clustering with an emulator directly trained on high-resolution $N$-body simulations. This comparison is particularly relevant for our purposes because the correlation dimension depends on a broad, top-hat-filtered range of Fourier modes rather than on a single value of $P(k,z)$.

The motivation is broader than this specific dark-energy model. In alternative scenarios such as $f(R)$ gravity or braneworld gravity, non-linear clustering can be altered by scale-dependent growth and by screening mechanisms that restore General Relativity in dense environments \cite{Dvali:2000hr,Hu:2007nk,Koyama_2009}. The modeling of the non-linear matter power spectrum in these theories has been extensively studied through simulations, perturbation theory, halo-model extensions, and reaction methods \cite{Oyaizu:2008tb,Schmidt_2009,Winther:2015wla,Cataneo:2018cic}. Although such modified-gravity models are not analyzed quantitatively here, they motivate our broader goal of understanding how non-linear corrections are propagated into the correlation dimension. In this sense, the $w_0w_a$CDM analysis presented in this paper should be regarded as a first step towards using $D_2$ as a geometrical probe of non-linear structure formation beyond $\Lambda$CDM.

The paper is organized as follows. In Sec.~\ref{Dq_param_section} we introduce the generalized fractal dimensions $D_q(r)$ and discuss in detail the connection between the correlation dimension $D_2(r)$ and the matter power spectrum $P(k)$. 
Beyond the descriptive presentation of the formalism presented in Sect.~\ref{sec:description}, we derive and analyze in Sect.~\ref{sec:approx} a simplified analytical approximation of the exact relation between $D_2(r)$ and $P(k)$, based on an effective cut-off representation of the top-hat filtering kernel. This approximation provides both a computationally efficient way to estimate $D_2$ and a useful interpretative tool to understand the range of Fourier modes that contributes to the scale dependence of the correlation dimension. In Sec.~\ref{NL_effects_sens} we study the impact of non-linear evolution on $D_2$ in the context of the $(w_0,w_a)$ dark energy model. Two complementary sensitivity analyzes are performed. First, we introduce a local sensitivity analysis, closely related to the Fisher-matrix approach, in which derivatives of the observable with respect to the cosmological parameters are evaluated around a fiducial model. This allows us to identify the scales and redshifts at which $D_2$, and in particular the ratio between its non-linear and linear predictions, responds most strongly to each parameter. Second, we performed a Sobol analysis in order to quantify the global sensitivity of $D_2$ to variations of the cosmological parameters across the full parameter space. Finally, Sec.~\ref{discussion} summarizes and discusses the main results, emphasizing their physical interpretation and implications for future applications of fractal observables in precision cosmology.

\section{Multifractal Dimensions in Cosmology}
\label{Dq_param_section}

The large-scale distribution of matter in the Universe is not spatially uniform at all scales. On small and intermediate scales, gravitational instability amplifies the primordial density
fluctuations and gives rise to the cosmic web, a complex network of clusters, filaments, walls and voids. On sufficiently large scales, however, the distribution is expected to approach statistical homogeneity, in agreement with the cosmological principle. Fractal and multifractal descriptors provide a natural geometrical language to quantify this transition.

The basic idea is to characterize how the amount of matter, or equivalently the number of
tracers, contained within a sphere of radius $r$ changes with the scale. For a perfectly
homogeneous distribution in three spatial dimensions, the enclosed number of objects scales
as $N(<r) \propto r^3$. A departure from this cubic scaling indicates clustering, and the
corresponding logarithmic slope defines an effective dimension. In cosmology, this effective
dimension is scale-dependent: it is lower than three on clustered scales and tends to three as the homogeneous regime is approached.

Multifractal dimensions generalize this idea by considering not only the mean scaling of the
distribution, but the scaling of different moments of the counts-in-spheres. In this way, they provide a hierarchy of scale-dependent observables which weight overdense and underdense regions differently. This is particularly useful in the cosmic web, where the matter distribution is highly inhomogeneous and the geometry cannot be fully characterized by a single number.

\subsection{General definition of fractal dimension $D_q(r)$}

Let us consider a discrete set of tracers, such as galaxies, halos or simulation particles, with mean number density $\bar{n}$. Around each tracer $i$, the number of neighbors
inside a sphere of radius $r$ is defined as
\be
N_i(<r)= 
\sum_{j\neq i}
\Theta\left(r-\left|\bm{x}_j-\bm{x}_i\right|\right),
\ee
where $\Theta$ is the Heaviside step function. The volume of the sphere is $V(r)=4\pi r^3/3$  
and the quantity $\bar{n}V(r)$ is the expected number of neighbors in a homogeneous Poisson
distribution with the same mean number density. Therefore, the ratio $N_i(<r)/(\bar{n}V(r))$  
measures the local excess or deficit of tracers around the point $i$ with respect to a
homogeneous reference distribution.   
The $q$-th order correlation sum is then defined as
\be
C_q(r) = \frac{1}{N}
\sum_{i=1}^{N}
\left[
\frac{N_i(<r)}{\bar{n}V(r)}
\right]^{q-1}\,,
\ee
and the associated generalized, or R\'enyi, dimension is
\be
D_q(r) = 3 + \frac{1}{q-1}
\frac{{\rm d}\ln C_q(r)}{{\rm d}\ln r}.
\ee
This definition is explicitly scale-dependent. Hence, $D_q(r)$ should not be interpreted as
a single asymptotic fractal dimension, but rather as an effective dimension measured at the
scale $r$. 

The parameter $q$ controls which regions of the density field dominate the statistic. For
$q>1$, over-dense regions contribute more strongly, because spheres with large neighbor
counts are weighted more heavily. For $q<1$ under-dense regions become more relevant.
The case $q=2$ is especially important because it is directly related to the two-point
correlation function and therefore connects the multifractal description with standard
large-scale structure statistics. 

In the homogeneous limit, the neighbor counts scale as $N_i(<r)\simeq \bar{n}V(r)\propto r^3$, and all generalized dimensions approach the ambient spatial dimension, $D_q(r)\longrightarrow 3$. 
On smaller scales, where clustering is relevant, one generally expects $D_q(r)<3$,  
with the precise value and scale dependence encoding the geometry of the tracer distribution.

\section{The correlation dimension $D_2(r)$ and the two-point correlation function}
\label{sec:description}

The case $q=2$ defines the correlation dimension. It is the most directly connected with the
usual two-point description of large-scale structure. For this value of the parameter, the correlation integral is
essentially proportional to the mean number of neighbors within a sphere of radius $r$, $N(<r)\propto \left\langle N_i(<r)\right\rangle$. 
For a statistically homogeneous and isotropic distribution, this mean count can be written in terms of the two-point correlation function $\xi(r)$ as
\be
N(<r) =
\bar{n}
\int_{V(r)}
\left[1+\xi(s)\right]{\rm d}^3s 
=
4\pi\bar{n}
\int_0^r
\left[1+\xi(s)\right]s^2{\rm d}s\,,
\ee 
where the last equality holds under the assumption of isotropy.  
It is useful to introduce the volume-averaged correlation function
\be
\bar{\xi}(r)
=
\frac{3}{r^3}
\int_0^r
\xi(s)s^2{\rm d}s\,,
\label{vacorrfunct}
\ee
and then the counts-in-spheres can be written as $N(<r) = \bar{n}V(r)\left[1+\bar{\xi}(r)\right]$. 
The correlation dimension is defined as the logarithmic derivative of the counts-in-spheres, that is,
\be
D_2(r)
= \frac{{\rm d}\ln N(<r)}{{\rm d}\ln r} 
= 3+ \frac{\rm d}{{\rm d}\ln r} \ln\left[1+\bar{\xi}(r)\right].
\label{eq:D2_xibar}
\ee

Equation~\eqref{eq:D2_xibar} explicitly shows that $D_2(r)$ measures the deviation from
the homogeneous scaling induced by the volume-averaged two-point correlation function. If
$\bar{\xi}(r)$ becomes negligible on large scales, then $D_2(r)\longrightarrow 3$, 
as expected for a homogeneous distribution. Conversely, when clustering is important, 
$\bar{\xi}(r)$ depends non-trivially on scale, and the second term in
Eq.~\eqref{eq:D2_xibar} shifts the correlation dimension below three. 

The previous expression also clarifies the connection with the usual power-law intuition. If, over a given range of scales, the two-point correlation function behaves approximately as $\xi(r)\propto r^{-\gamma}$, 
then its volume average has the same scaling, $\bar{\xi}(r)\propto r^{-\gamma}$. In the 
strongly clustered regime, where $\bar{\xi}(r)\gg 1$, Eq.~\eqref{eq:D2_xibar} gives $D_2(r)\simeq 3-\gamma$.  
Thus, the correlation dimension translates the slope of the two-point correlation function
into a geometrical measure of clustering. This provides the first bridge between the
multifractal description and the standard statistical observables used in cosmology.

The approach of $D_2(r)$ to the homogeneous value provides a natural
operational definition of the homogeneity scale $R_{\rm H}$. Since in a finite survey
the limit $D_2(r)\rightarrow 3$ is reached only asymptotically and is affected
by statistical fluctuations, one usually defines $R_{\rm H}$ as the smallest
scale above which the correlation dimension is sufficiently close to homogeneity. Following the standard one-percent criterion, we define
\be
D_2(R_{\rm H}) = 2.97\,,
\label{eq:RH_def}
\ee
or, more precisely, $R_{\rm H}$ is the first radius such that
$D_2(r)\geq D_2(R_{\rm H})$ for all $r\geq R_{\rm H}$ within the range probed by the
analysis. The value $2.97$ is therefore not arbitrary: it corresponds to
requiring the measured scaling of the counts-in-spheres to be within
$1\%$ of the homogeneous expectation, $D_2=3$. This provides a practical
and survey-independent way of assigning a finite transition scale to a
gradual approach to homogeneity \cite{Hogg:2004vw, Ntelis:2017nrj, Scrimgeour:2012wt}.

\subsection{Fourier-space expression and the role of the top-hat window}

The previous expression for the correlation dimension can be rewritten in Fourier space
by using the relation between the two-point correlation function and the matter power
spectrum. For a statistically homogeneous and isotropic density field, the two-point
correlation function is
\be
\xi(r) = \frac{1}{2\pi^2}
\int_0^\infty k^2 P(k)
\frac{\sin(kr)}{kr} \,{\rm d}k\,.
\ee
The volume-averaged correlation function entering the counts-in-spheres follows Eq.~\eqref{vacorrfunct}, so we replace the Fourier expression of $\xi(s)$ and, exchanging the order of integration, we obtain
\be
\bar{\xi}(r)
=
\frac{1}{2\pi^2}
\int_0^\infty
k^2P(k)W_{\rm TH}(kr)\,{\rm d}k\,,
\label{eq:xibar_pk}
\ee
where $W_{\rm TH}(x) = 3 \left(\sin x-x\cos x\right)/x^3$ is the Fourier transform of a spherical real-space top-hat window. Therefore, the correlation dimension can be computed directly from the matter power
spectrum as
\be
D_2(r, z) = 3 +
\frac{\rm d}{{\rm d}\ln r}
\ln \left[ 1+ \frac{1}{2\pi^2}
\int_0^\infty
k^2P(k, z)W_{\rm TH}(kr)\,{\rm d}k
\right]\,.
\label{eq:D2_pk}
\ee
This expression provides the connection between the geometrical description of clustering
through $D_2(r,z)$ and the standard Fourier-space description through $P(k,z)$, explicitly accounting for their redshift dependence. 

It is often convenient to introduce the dimensionless matter power spectrum $\Delta^2(k) = k^3P(k)/(2\pi^2)$. 
Then, since ${\rm d}k/k={\rm d}\ln k$, Eq.~\eqref{eq:xibar_pk} can be written as
\be
\bar{\xi}(r)
=
\int_0^\infty
\Delta^2(k)W_{\rm TH}(kr)\,{\rm d}\ln k\,.
\ee
Defining $F(r) \equiv 1+\bar{\xi}(r)$, the exact correlation dimension becomes
\be
D_2(r) = 3+
\frac{{\rm d}\ln F(r)}{{\rm d}\ln r} = 3 + \frac{r \, F'(r)}{F(r)}\,,
\label{eq:D2_F}
\ee
where,
\be 
r \, F'(r) = r \int_0^\infty
\Delta^2(k)\, W'_{\rm TH}(kr)\,k\,{\rm d}\ln k\,,
\ee
being $W'_{\rm TH}(x) = {\rm d}W_{\rm TH} / {\rm d}x$.

The role of the window function is central. The quantity $D_2(r)$ does not depend on a single Fourier mode of the power spectrum, but rather receives contributions from a weighted range of modes. For a fixed radius $r$, the argument of the window is $x=kr$; this implies that the modes with $kr\ll 1$ have wavelengths much larger than the sphere. In this limit, $W_{\rm TH}(x) = 1-x^2/10+\mathcal{O}(x^4)$, 
so these modes contribute almost coherently to the volume-averaged correlation
function. However, their contribution to the logarithmic derivative is suppressed because $xW_{\rm TH}'(x) = -x^2/5 +\mathcal{O}(x^4)$. 
Therefore, very long-wavelength modes affect the normalization of $F(r)$ but contribute
weakly to the scale variation of $F(r)$. 
 On the other hand, modes with $kr\gg 1$ have wavelengths much smaller than the
sphere. In this regime, the top-hat window becomes oscillatory and decays in amplitude, $W_{\rm TH}(x) \sim -3\cos x/x^2 + 3\sin x/x^3$. 
As a consequence, high-$k$ modes are strongly suppressed in the integral for
$\bar{\xi}(r)$, apart from residual oscillatory contributions that tend to cancel out. The
dominant contribution to the scale dependence of $D_2(r)$ therefore comes from modes
around the transition region $k\sim 1/r$. 
This relation should be understood as an order-of-magnitude correspondence rather than
as a sharp equality. Since $W_{\rm TH}(kr)$ is not a compact window in the Fourier space,
a given radius $r$ receives contributions from a finite band of wavenumbers. Nevertheless,
the oscillatory structure of $W_{\rm TH}$ implies that the integral is effectively dominated
by modes with $k$ of order a few times $1/r$, while modes far above this range are
suppressed by cancellations. This is the physical reason why $D_2(r)$ can be regarded as
a smoothed, scale-dependent probe of the shape of $\Delta^2(k)$. 

The Fourier-space expression also makes explicit how cosmological information enters the correlation dimension. A change in the cosmological model modifies the matter power
spectrum, and this modification is propagated to $D_2(r,z)$ through the non-local integral kernel
defined by $W_{\rm TH}(kr)$. Therefore, $D_2(r)$ is not simply a reparameterization of
$P(k)$ at $k=1/r$, but a filtered observable that mixes neighboring Fourier modes.

The discussion above also motivates the construction of simplified approximations. Since
the window $W_{\rm TH}(kr)$ behaves approximately as a low-pass filter, one may try to
replace the smooth oscillatory kernel with an effective sharp cut-off in Fourier space. This
leads to an approximate expression in which the relevant accumulated power at radius
$r$ is represented by an integral of $\Delta^2(k)$ up to a calibrated scale $k_{\rm cut}$. 
The derivation and global calibration of this cut-off approximation will be discussed in the next
section. 

\subsection{Nonlinear correlation dimension} 

The non-linear corrections to the correlation dimension 
are naturally computed from a nonlinear matter power spectrum, $P^{\rm nl}(k,z)$. In order to explore how nonlinearities propagate from one observable to the other, we work with two particular approaches for $P^{\rm nl}(k,z)$ using (i) the Euclid Emulator 2 (EE2) \cite{Euclid:2018mlb,Euclid:2020rfv} and (ii) the augmented halo model HMCode2020 \cite{Mead:2015yca,Mead:2020vgs}. While EE2 provides a high-precision emulator calibrated directly on $N$-body simulations, HMCode2020 relies on a semi-analytic halo-model description calibrated to simulations. Comparing the resulting $D_2^{\rm nl}(r,z)$ predictions therefore provides an independent assessment of the robustness of the correlation dimension to the modeling of nonlinear structure formation. In particular, we can follow the propagation of differences from $P^{\rm nl}$ to the correlation function $\xi$ and finally to $D_2^{\rm nl}$, since the Fourier transform and logarithmic derivative can modify or amplify differences between nonlinear prescriptions. The comparison is expected to be most informative on the intermediate and small scales, $r\sim1-10$\,Mpc/$h$, where nonlinear effects become important. 

Furthermore, EE2 and HMCode2020 are built for computing not only the standard cosmological model $\Lambda$CDM, but also the extended $w_0w_a$CDM model for dynamical dark energy. As mentioned above, this is an interesting parameterization in light of recent BAO measurements from DESI DR2, which have shown a preference for this model over a cosmological constant scenario when combined with CMB or CMB+SNe data \cite{DESIdr2}. We are able then to compute the effects of nonlinear corrections in these two cosmologies and explore possible distinctive features on the observables. 

\begin{figure*}
\centering
\includegraphics[width=0.49\linewidth]{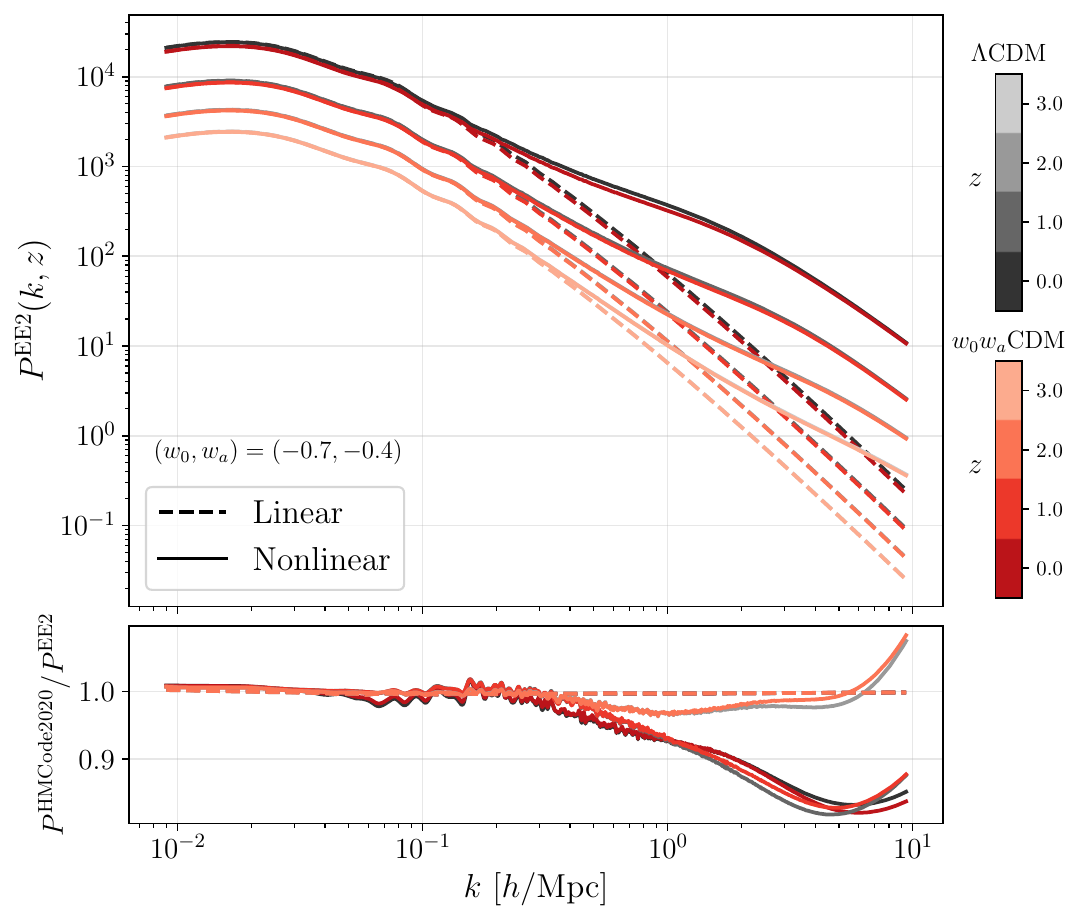}
\includegraphics[width=0.49\linewidth]{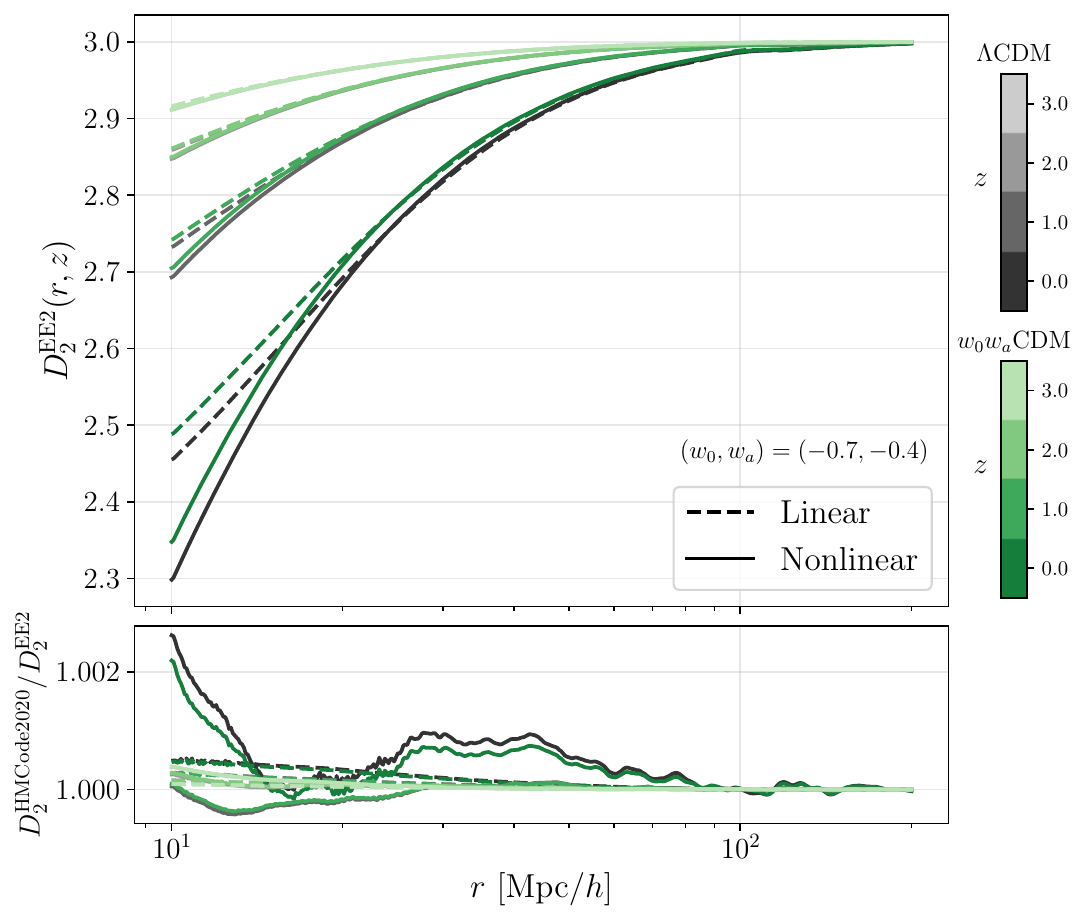}
\caption{General behavior of the power spectrum $P(k,z)$ (left) and the correlation dimension $D_2(r,z)$ (right) as a function of scale for $\Lambda$CDM and $w_0w_a$CDM cosmologies at different redshifts (color gradient). Solid and dashed lines correspond to the linear and nonlinear predictions, respectively. The comparison illustrates the effect of nonlinear evolution on the corelation dimension across scales and redshifts. The bottom panels compare the predictions from EE2 and HMCode2020, showing that $D_2$ is robust  against the choice of nonlinear prescription. 
}
\label{fig:comparison}
\end{figure*}

The general behavior of both $P^{\rm nl}(k)$ and $D_2^{\rm nl}(r,z)$ is shown in Fig.~\ref{fig:comparison} at different redshifts for the $\Lambda$CDM and $w_0w_a$CDM models, considering two fiducial cosmologies with parameters 
$A_s=2.105\times10^{-9}$, $n_s=0.9665$, $\Omega_b=0.049$, $\Omega_m=0.3096$, $h=0.67$, $m_\nu=0.06$, and  $(w_0,w_a)_{\Lambda{\rm CDM}} =(-1,0)$ and $(w_0,w_a)_{w_0w_a{\rm CDM}}=(-0.7,-0.4)$. As expected, the differences between the two cosmologies on small scales become significant only at very low redshift, where the dynamical dark energy component departs most strongly from a cosmological-constant background. This corresponds to a relative increase of around $\sim 0.06\%$ with respect to their linear predictions.

Moreover, while nonlinear effects enhance the matter power spectra on small scales, they produce  a suppression of the correlation dimension towards smaller scales. Indeed, this behavior is expected since $D_2<3$ quantifies deviations from homogeneity. It is also notable that $D_2$ is robust  against the choice of nonlinear prescription, with the differences between the EE2 and HMCode2020 predictions remaining at the level of $10^{-3}$ for both cosmologies considered. 

\section{Cut-off approximation for $D_2$}
\label{sec:approx}

The exact expression for the correlation dimension $D_2(r)$ involves the full Fourier-space top-hat window $W_{\rm TH}(x)$, whose oscillatory form makes its contribution difficult to analyze in a simple way. Moreover, the expression requires differentiating an improper integral over the full range $0<k<\infty$. 
For this reason, it is useful to derive an approximate expression that simplifies both the role of the window function and the differentiation of the integral. The goal is to obtain a more transparent relation between the shape of the power spectrum and the resulting behavior of the correlation dimension.

As discussed above, the top-hat window behaves qualitatively as a smooth low-pass
filter in a Fourier space. For a fixed radius $r$, the modes with $kr\ll 1$ contribute almost coherently to the averaged correlation function, while the modes with $kr\gg 1$ are
suppressed by the oscillatory decay of $W_{\rm TH}(kr)$. This motivates replacing the
smooth oscillatory kernel by an effective sharp cut-off in Fourier space. We define
\be
k_{\rm cut}(r) = \frac{\alpha}{r},
\label{eq:kcut_definition}
\ee
where $\alpha$ is a dimensionless parameter. The cut-off approximation consists of the
replacement
\be
W_{\rm TH}(kr)
\quad \longrightarrow \quad
\Theta\!\left(k_{\rm cut}(r)-k\right)\,.
\ee
Under this approximation, the exact function $F(r)$ is replaced by
\be
F_{\rm cut}(r) = 1+
\int_0^{k_{\rm cut}(r)}
\Delta^2(k)\,{\rm d}\ln k\,.
\label{eq:Fcut_definition}
\ee
That is, the volume-averaged correlation contribution is approximated by the cumulative
dimensionless power up to the effective scale $k_{\rm cut}(r)$. 

The corresponding cut-off expression for the correlation dimension is obtained by
taking the logarithmic derivative of Eq.~\eqref{eq:Fcut_definition}. Since the upper
limit of integration depends on $r$, we use
\be
\frac{\rm d}{{\rm d}\ln r}
\int_0^{k_{\rm cut}(r)}
\Delta^2(k)\,{\rm d}\ln k =
\Delta^2(k_{\rm cut})
\frac{{\rm d}\ln k_{\rm cut}}{{\rm  d}\ln r}.
\ee
From Eq.~\eqref{eq:kcut_definition}, $\ln k_{\rm cut} = \ln\alpha-\ln r$ 
and therefore ${\rm d}\ln k_{\rm cut}/{\rm d}\ln r = -1$.
Hence, ${\rm d}F_{\rm cut}/{\rm d}\ln r = -\Delta^2(k_{\rm cut})$, 
and substituting this into the definition of the correlation dimension gives
\be
D_{2,{\rm cut}}(r;\alpha) = 3+ 
\frac{1}{F_{\rm cut}(r)}
\frac{{\rm d}F_{\rm cut}(r)}{{\rm d}\ln r} = 3-
\frac{ \Delta^2(\alpha/r)
}{\displaystyle
1+ \int_0^{\alpha/r)}
\Delta^2(k)\,{\rm d}\ln k
}\,.
\label{eq:D2_cutoff}
\ee

This formula has a simple physical interpretation. The denominator measures the
cumulative clustering contribution from all modes with a wavenumber below the effective
cut-off scale. The numerator measures the amount of dimensionless power precisely at
the moving limit $k_{\rm cut}(r)$. As the radius $r$ increases, the cut-off scale
decreases and the accumulated contribution changes according to the power that
crosses the limit. The minus sign in Eq.~\eqref{eq:D2_cutoff} is a direct consequence of
the fact that increasing $r$ lowers $k_{\rm cut}$. 

The approximation can be viewed as replacing the full non-local functional
\be
D_{2,{\rm exact}}(r) = 3+
\frac{\rm d}{{\rm d}\ln r}
\ln \left[ 1+ \int_0^\infty
\Delta^2(k)W_{\rm TH}(kr)\,{\rm d}\ln k
\right]
\ee
by the simpler functional given by Eq.~\eqref{eq:D2_cutoff}.   
Both quantities depend on the same input function $\Delta^2(k)$, but encode the
scale dependence in different ways. The exact expression uses the full oscillatory
top-hat kernel, whereas the cut-off expression compresses its effect into a single moving
Fourier scale $k_{\rm cut}=\alpha/r$.  

The parameter $\alpha$ therefore controls which Fourier modes are assigned to a given
real-space radius. If $\alpha$ is too small, the approximation excludes modes that still
contribute significantly to the exact top-hat-filtered integral. If $\alpha$ is too large, it includes modes whose contributions are largely canceled by the oscillatory behavior of
the exact window. Thus, $\alpha$ cannot be fixed purely by dimensional analysis; it must
be determined by comparing the cut-off functional with the exact one. This calibration
will be discussed in the next subsection.

\subsection{Global calibration of the cut-off parameter $\alpha$}

The cut-off approximation derived in the previous subsection depends on a single dimensionless parameter, $\alpha$, through the effective relation of $k_{\rm cut}(r)$. This parameter determines which Fourier scale is associated with a given real-space
radius. Since the exact expression involves the full top-hat window $W_{\rm TH}(kr)$,
whereas the cut-off approximation replaces it by a sharp cumulative integral, $\alpha$
cannot be fixed by dimensional arguments alone. Instead, it must be calibrated by
requiring the approximate functional to reproduce the exact one as accurately as possible. 

The exact and approximate expressions can be viewed as two different mappings from
the input dimensionless power spectrum $\Delta^2(k)$ to the output correlation
dimension:
\be
D^{\rm exact}_2(r) = {\cal F}_{\rm exact}\!\left[\Delta^2\right](r),
\ee
and
\be
D^{\rm cut}_2(r;\alpha) = 
{\cal F}_{\rm cut}\!\left[\Delta^2;\alpha\right](r).
\ee
The objective of the calibration is to determine a single value of $\alpha$ that minimizes
the distance between these two functionals over a broad region of cosmological parameter
space. In this way, the cut-off approximation is not tuned to one fiducial cosmology only,
but is instead optimized to remain accurate when the underlying cosmological model is
varied. 

For this purpose, a set of cosmological models was generated by randomly sampling the
parameter vector within the ranges listed in Table~\ref{tab:alpha_sampling_ranges}. For
each cosmology, both the linear and the non-linear matter power spectra were computed.
The corresponding dimensionless spectra $\Delta^2(k)$ were then used to evaluate both
the exact and cut-off approximated expressions for $D_2$. 

\begin{table*}
\centering
\setlength{\tabcolsep}{9pt}
\begin{tabular}{lccccccccccc}
\hline\hline
   & $z$ & $H_0$ & $\Omega_bh^2$ & $\Omega_ch^2$ & $\Omega_k$ 
 & $\tau$ & $A_s$ & $n_s$ & $w_0$ & $w_a$ & $m_\nu$ \\
\hline
Minimum & $0.0$ & $61.0$ & $0.0196$ & $0.117$ & $-0.01$ 
& $0.045$ & $1.7\times 10^{-9}$ & $0.92$ & $-1.3$ & $-0.7$ & $0.0$ \\
Maximum & $3.0$ & $73.0$ & $0.0294$ & $0.196$ & $0.01$ & $0.070$ & $2.5\times 10^{-9}$ & $1.00$ & $-0.7$ & $0.5$ & $0.15$ \\
\hline\hline
\end{tabular}
\caption{ 
Parameter ranges used in the random sampling of cosmological models for the
global calibration of the cut-off approximation. The scan explores variations in
redshift, background expansion, baryon and cold dark matter densities, spatial
curvature, optical depth, primordial amplitude and tilt, dark-energy dynamics and
neutrino mass. We have fixed $N_{\rm massive} = 1$ and $N_{\rm eff} = 3.044$.
}
\label{tab:alpha_sampling_ranges}
\end{table*}

For a given cosmological model $m$ and a fixed value of $\alpha$, the exact and
approximate curves are compared in a discrete set of radii $\{r_i\}_{i=1}^{N_r}$.
The comparison is performed separately for the linear and non-linear spectra. Thus, for
the linear case, we consider the two vectors $D^{\rm exact,(m)}_{2,{\rm lin}}(r_i)$ and $D^{\rm cut,(m)}_{2,{\rm lin}}(r_i;\alpha)$,  
and analogously for the non-linear case, $D^{\rm exact,(m)}_{2,{\rm nl}}(r_i)$ and $D^{\rm cut,(m)}_{2,{\rm nl}}(r_i;\alpha)$.  
The error between two curves is quantified using a Normalized Mean Squared Error (NMSE), defined as 
\be
{\rm NMSE}(y_{\rm true},y_{\rm pred})
=
\frac{
\frac{1}{N_r}\sum_{i=1}^{N_r}
\left(y_{{\rm pred},i}-y_{{\rm true},i}\right)^2
}{
\frac{1}{N_r}\sum_{i=1}^{N_r}
\left(y_{{\rm true},i}-\bar{y}_{\rm true}\right)^2
+\epsilon
}\,,
\label{eq:NMSE_definition}
\ee
where $y_{\rm true}$ and $y_{\rm pred}$ are the true and predicted curves, respectively, 
$\bar{y}_{\rm true} = \left(\sum_{i=1}^{N_r}y_{{\rm true},i}\right)\big/N_r $, 
and $\epsilon$ is a small regulator introduced to avoid numerical instabilities if the
variance of the exact curve becomes very small. 

For each cosmological model $m$, the linear and non-linear errors are therefore   
\be
E^{(m)}_{\rm lin}(\alpha) = {\rm NMSE} \left(
D^{\rm exact,(m)}_{2,{\rm lin}},
D^{\rm cut,(m)}_{2,{\rm lin}}(\alpha) \right) 
\quad {\rm and} \quad  
E^{(m)}_{\rm nl}(\alpha)= {\rm NMSE}
\left(
D^{\rm exact,(m)}_{2,{\rm nl}},
D^{\rm cut,(m)}_{2,{\rm nl}}(\alpha)
\right)\,.
\ee
The total error for that model is defined as a weighted sum: 
$E^{(m)}_{\rm model}(\alpha) = w_{\rm lin}E^{(m)}_{\rm lin}(\alpha)
+ w_{\rm nl}E^{(m)}_{\rm nl}(\alpha)$. 
In the calibration presented here, both contributions are weighted equally, so that $w_{\rm lin}=w_{\rm nl}=1$.  
The global objective function is obtained by averaging the model error over the full
ensemble of sampled cosmologies:
\be
E_{\rm global}(\alpha)
=
\frac{1}{N_{\rm models}}
\sum_{m=1}^{N_{\rm models}}
E^{(m)}_{\rm model}(\alpha)\,,
\label{eq:Eglobal_alpha}
\ee
and the optimal cut-off parameter is then defined by 
\be
\alpha_{\rm best} = \arg\min_\alpha E_{\rm global}(\alpha)\,.
\label{eq:alpha_best_definition}
\ee

\begin{figure*}
\centering
\includegraphics[width=0.48\linewidth]{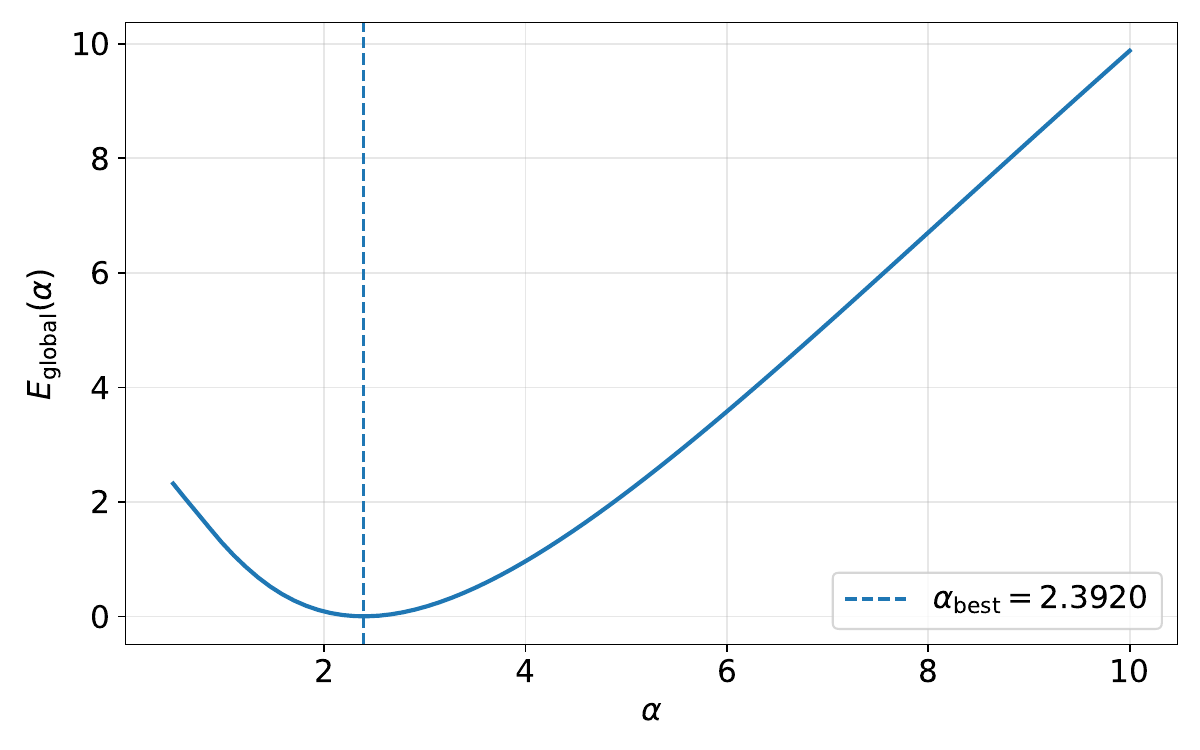}
\includegraphics[width=0.48\linewidth]{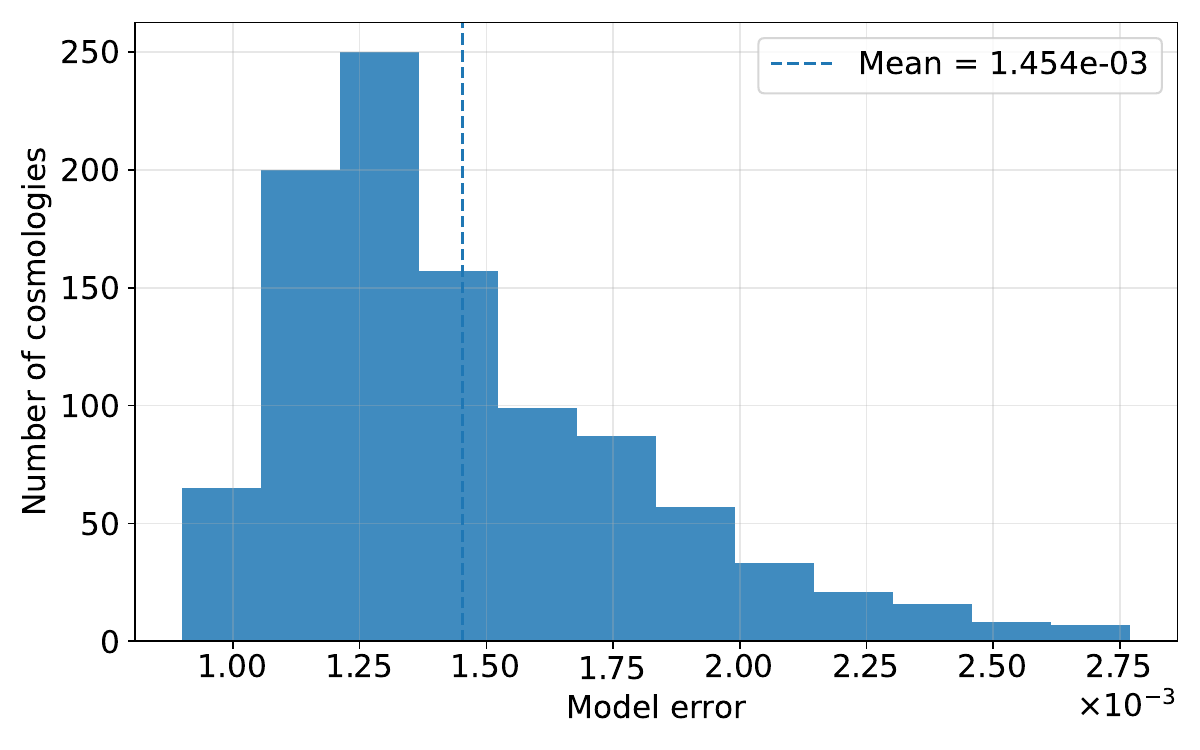}
\caption{{\bf Left.}
Global objective function $E_{\rm global}(\alpha)$ as a function of the cut-off
parameter $\alpha$. The curve is obtained by averaging the model error over the full
set of sampled cosmologies, including both the linear and non-linear contributions.
The vertical dashed line marks the best-fit value $\alpha_{\rm best}=2.392$. The
existence of a sharp minimum shows that the top-hat window can be effectively
represented, in the present approximation, by a moving cut-off scale
$k_{\rm cut}=\alpha/r$ with a well-defined value of $\alpha$. {\bf Right.} Histogram of the individual model errors
$E^{(m)}_{\rm model}(\alpha_{\rm best})$ evaluated at the global best-fit value
$\alpha_{\rm best}=2.392$. The distribution, with mean value
$1.454\times 10^{-3}$, indicates that the globally calibrated cut-off approximation
does not work only for a small subset of models, but provides a robust description
across the full cosmological sample.
}
\label{fig:alpha_global_error_histogram}
\end{figure*}

The minimization was carried out in two steps. First, a coarse scan was performed on a regular grid of $\alpha$ values to locate the approximate position of the minimum. Then, a local refinement around the best grid value was performed using a bounded scalar minimization. This procedure yields $\alpha_{\rm best}=2.392$ for the calibration based on $N_{\rm models}=1000$ random cosmologies.   
The left panel of Fig.~\ref{fig:alpha_global_error_histogram} shows the global objective function
$E_{\rm global}(\alpha)$. The curve displays a well-defined minimum, indicating that
there is a preferred effective cut-off scale that best reproduces the action of the full
top-hat window over the sampled cosmological domain. The vertical dashed line indicates the best-fit value $\alpha_{\rm best}$. 
 The distribution of individual model errors evaluated at $\alpha_{\rm best}$ is shown in the right panel of Fig.~\ref{fig:alpha_global_error_histogram}. This histogram measures how uniformly the single global value of $\alpha$ performs in the cosmological ensemble. The distribution is asymmetric, with a tail that extends towards errors larger than the mean value $\langle E^{(m)}_{\rm model}(\alpha_{\rm best})\rangle = 1.454\times 10^{-3}$, 
showing that the same cut-off parameter provides a stable approximation throughout the
sampled parameter space. 

\begin{figure*}[t]
\centering
\includegraphics[width=0.92\textwidth]{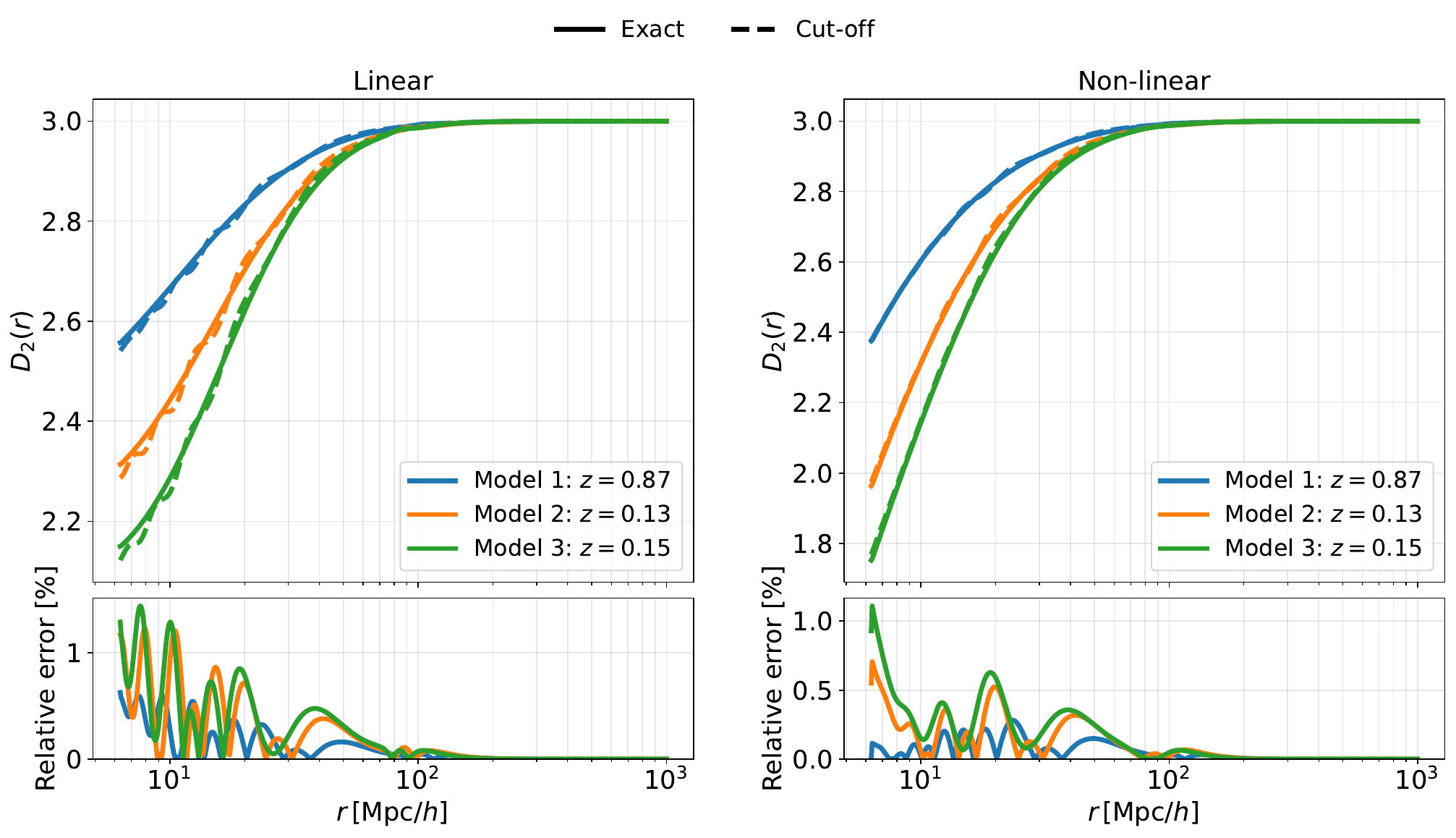}
\caption{
Comparison between the exact correlation dimension and the cut-off approximation
for three representative cosmological models. The left panel shows the linear case
and the right panel the non-linear case. Solid curves correspond to the exact
calculation using the full Fourier-space top-hat window, while dashed curves
represent the cut-off approximation calibrated with
$\alpha_{\rm best}=2.392$. The agreement is very good across the range of radii,
showing that the effective cut-off captures the dominant scale dependence of the
top-hat filtered integral. Relative percentage error between the exact correlation dimension and the cut-off
approximation are also shown. The error is largest at small radii, where the sharp cut-off is a more idealized representation of the oscillatory top-hat kernel, but it remains at the percent level in the examples shown. At large radii the error decreases rapidly, indicating that the calibrated cut-off approximation reproduces very accurately the approach to homogeneity.
}
\label{fig:D2_models_cutoff}
\end{figure*}

To illustrate the performance of the approximation at the level of the correlation
dimension itself, Fig.~\ref{fig:D2_models_cutoff} compares the exact and cut-off curves
for three representative cosmological models selected from the full ensemble. The left
panel corresponds to the linear matter power spectrum, while the right panel corresponds
to the non-linear spectrum. In both cases, the cut-off approximation follows closely the
exact result over the full range of radii shown. The comparison also shows the expected
approach to homogeneity, with $D_2(r)\rightarrow 3$ at large radii. In the non-linear case, the curves reach lower values of $D_2(r)$ at small
and intermediate scales, reflecting the enhancement of clustering induced by non-linear
evolution. 

The same comparison can be quantified through the relative percentage error,
\be
\epsilon_{\rm rel}(r) = 100\,
\left|\frac{
D^{\rm cut}_2(r;\alpha_{\rm best}) - D^{\rm exact}_2(r)
}{D^{\rm exact}_2(r)}
\right|.
\ee
Figure~\ref{fig:D2_models_cutoff} shows this error for the same three
representative cosmologies. The error remains at the percent level throughout the
range shown. It is largest on the smallest scales, where the oscillatory structure of the
exact top-hat window and the detailed scale dependence of $\Delta^2(k)$ are more
difficult to compress into a single sharp boundary. At larger radii, the relative error
rapidly decreases as the curves approach the homogeneous limit. 

As a robustness test, the same calibration procedure was repeated using an independent
sample of $300$ random cosmologies drawn from the same parameter ranges. This second
analysis gives $\alpha_{\rm best}=2.392$, which differs from the value of $N_{\mathrm{models}} = 1000$ only in the fourth
decimal place. This agreement indicates that the fitted value of $\alpha$ is not an
accidental feature of a particular random realization of the cosmological sample, but a
stable property of the approximation over the parameter domain considered here.

In summary, the globally calibrated value $\alpha_{\rm best}\simeq 2.392$ provides an effective mapping between a real-space radius $r$ and the Fourier cut-off
scale $k_{\rm cut}$. 
With this value, the cut-off approximation offers a compact
and analytically transparent representation of the exact top-hat filtered expression for
$D_2(r)$, while preserving high accuracy for both linear and non-linear matter power
spectra across a broad family of cosmological models.

\section{Global and local sensitivities of on cosmological parameters}
\label{NL_effects_sens}

In this Section, we investigate the local and global sensitivities of the correlation dimension to the parameters of the underlying cosmological model. These local and global approaches are complementary. While a local derivative-based analysis typically evaluates how observable changes under a small perturbation of a parameter $\theta_\alpha$ around a fiducial cosmology, the global Sobol analysis identifies which cosmological parameters contribute the most to the variance of $D_2(r,z)$ across the entire cosmological parameter space. This distinction is particularly relevant for our problem because the sensitivity can change with both scale $r$ and redshift $z$, while non-linear parameter interactions may also become important. 

\subsection{Local sensitivity of $D_2(r, z)$ and parameter correlation in $w_0w_a$CDM with massive neutrinos}

We analyze the dependence of the correlation dimension $D_2(r, z)$ on the cosmological parameters in a $w_0w_a$CDM model with massive neutrinos. The analysis is carried out for both linear and nonlinear cases, employing \texttt{HMcode2020} modeling for the latter. The goal is, on the one hand, to study how the values of $D_2(r, z)$ respond to variations in the model parameters and, on the other hand, to determine to what extent the different parametric dependencies are degenerate with each other.

We consider a fiducial $w_0w_a$CDM model with massive neutrinos, characterized by the set of parameters
\begin{equation}
\boldsymbol{\theta} =
\{H_0,\Omega_b h^2,\Omega_c h^2,A_s,n_s,w_0,w_a,m_\nu\}.
\end{equation}
The fiducial values adopted in the sensitivity analysis are
\begin{equation}
\boldsymbol{\theta}_{\rm fid} =
\{67.66,\;0.02242,\;0.11933,\;2.105\times10^{-9},\;0.9665,\;-1,\;0,\;0.06\,{\rm eV}\},
\end{equation}
where $H_0$ is given in units of ${\rm km/s/Mpc}$, $m_\nu$
denotes the sum of neutrino masses and we fix $\Omega_k=0$. The dark-energy equation of state is
parameterized as $w(a)=w_0+(1-a)w_a$ and then the fiducial model fits the $\Lambda$CDM cosmology. From this model, the matter power spectrum $P(k, z)$ is computed for both the linear and nonlinear cases, 
and then form them the linear and nonlinear correlation dimensions $D_2(r, z)$ are obtained.

In order to quantify the response of $D_2(r, z)$ to variations in the model parameters, we define the local sensitivity with respect to the parameter $\theta_\alpha$ as
\begin{equation}\label{loc_sens}
S_\alpha(r, z) = \frac{|\theta_\alpha|}{D_2(r, z)} \frac{\partial D_2(r, z)}{\partial \theta_\alpha}.
\end{equation}
In the case where $\theta_\alpha^{\rm fid} = 0$, we replace $\theta_\alpha$ as $1$ in Eq.~\eqref{loc_sens}. In practice, this derivative is evaluated numerically using a centered difference,
\begin{equation}
\frac{\partial D_2(r, z)}{\partial \theta_\alpha} \approx \frac{D_2(r, z; \theta_\alpha + h_\alpha) - D_2(r, z; \theta_\alpha - h_\alpha)}{2 h_\alpha},
\end{equation}
where the increment is taken as $h_\alpha = \varepsilon \, \theta_\alpha^{\rm fid}$, with $\varepsilon = 0.02$, 
and in the case where $\theta_\alpha^{\rm fid} = 0$, we adopt $h_\alpha = \varepsilon$. This quantity allows us to identify whether an increase in a parameter increases or decreases $D_2(r, z)$, as well as the scale and redshift at which the observable is most sensitive. We also compute this local sensitivity for the homogeneity scale $R_{\rm H}$ as a function of $z$.

To study the degeneracy between parameters, we construct, for each parameter $\theta_\alpha$, a one-dimensional vector $v_\alpha \equiv S_\alpha(r, z)$ from all sensitivity values on the radii and the redshift grid. 
 We then calculate the angular similarity between pairs of parameters using $\cos(\theta_{\alpha\beta}) = (v_\alpha \cdot v_\beta)/(\|v_\alpha\|\,\|v_\beta\|)$.
 Values close to 1 indicate that two parameters produce very similar responses in $D_2(r, z)$, suggesting strong degeneracy. Values near $-1$ correspond to responses of similar shape but with opposite sign, while values close to 0 indicate more differentiated responses.

\begin{figure*}
\centering
\includegraphics[width=1.0\textwidth]{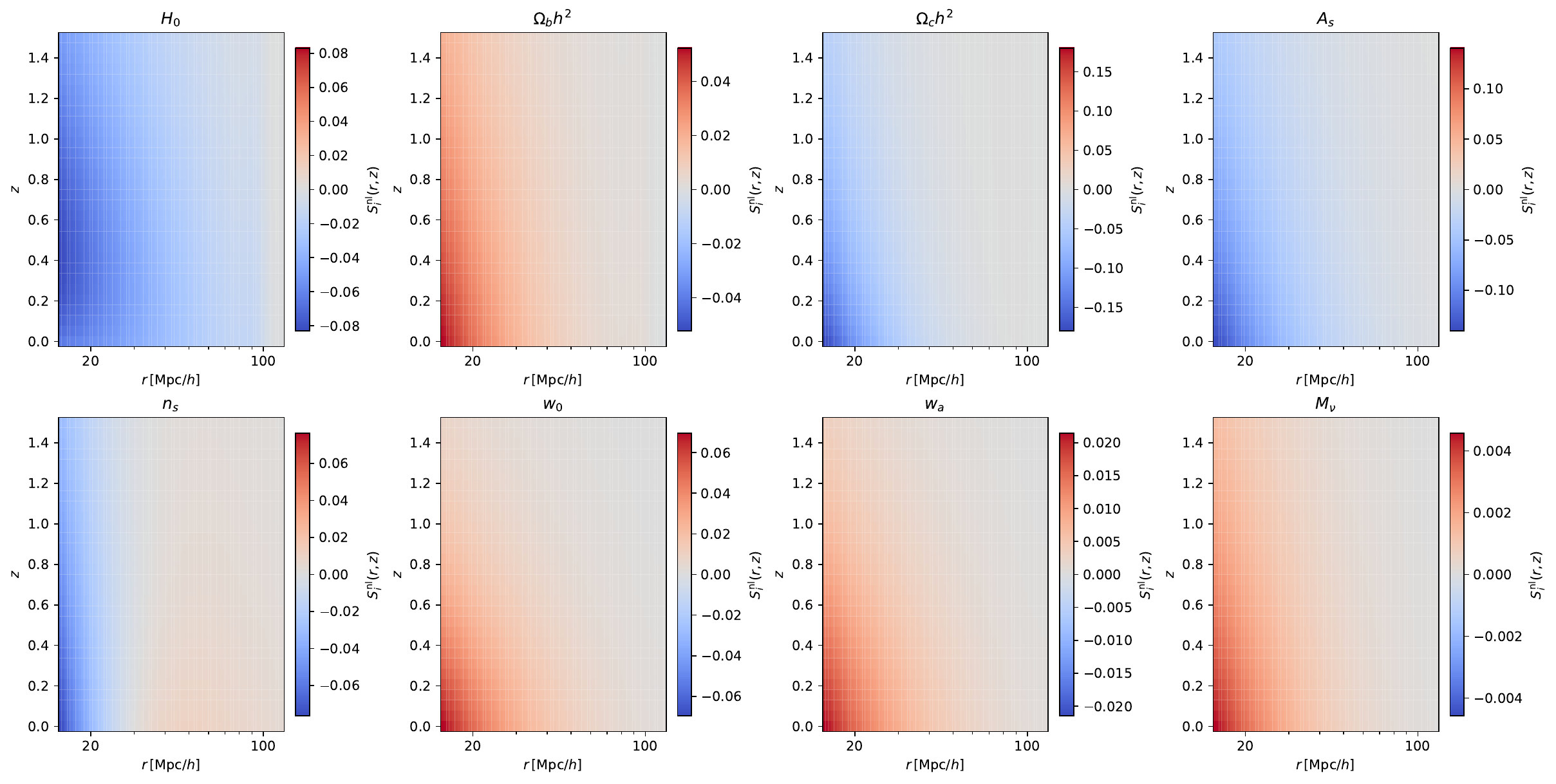}
\caption{Local sensitivities of the nonlinear correlation dimension $D_2^{\rm nl}(r,z)$ with respect to the cosmological parameters. The strongest responses appear at small scales and low redshift. In general, increasing $\Omega_b h^2$, $w_a$ and $m_\nu$ tends to increase $D_2^{\rm nl}$, whereas $\Omega_c h^2$, $A_s$, $n_s$ and $w_0$ tend to reduce it over the regions of largest sensitivity.
}
\label{fig:nonlinear_loc_sens}
\end{figure*}

Figure~\ref{fig:nonlinear_loc_sens} shows the local sensitivities of the
nonlinear correlation dimension, $D_2^{\rm nl}(r,z)$, for the set of
cosmological parameters considered in this analysis. Each panel represents the
dimensionless response $S_\alpha^{\rm nl}(r,z)$ on the two-dimensional grid of
scales and redshifts. The sensitivity is not uniform across the parameter space:
the largest responses are concentrated at relatively small scales and low
redshift, where nonlinear clustering is more relevant. At larger scales, close
to the approach to homogeneity, the response becomes progressively weaker.  
The different parameters also show distinct signs and amplitudes. Variations in
$\Omega_b h^2$, $w_a$ and $m_\nu$ tend to increase $D_2^{\rm nl}$ in the
regions where the sensitivity is largest, whereas $\Omega_c h^2$, $A_s$, $n_s$
and $w_0$ mainly produce the opposite effect. Among the parameters shown,
$\Omega_c h^2$, $A_s$, $n_s$ and $w_0$ display some of the largest amplitudes,
while the sensitivity to $m_\nu$ is comparatively smaller.

\begin{figure*}
\centering
\includegraphics[width=1.0\textwidth]{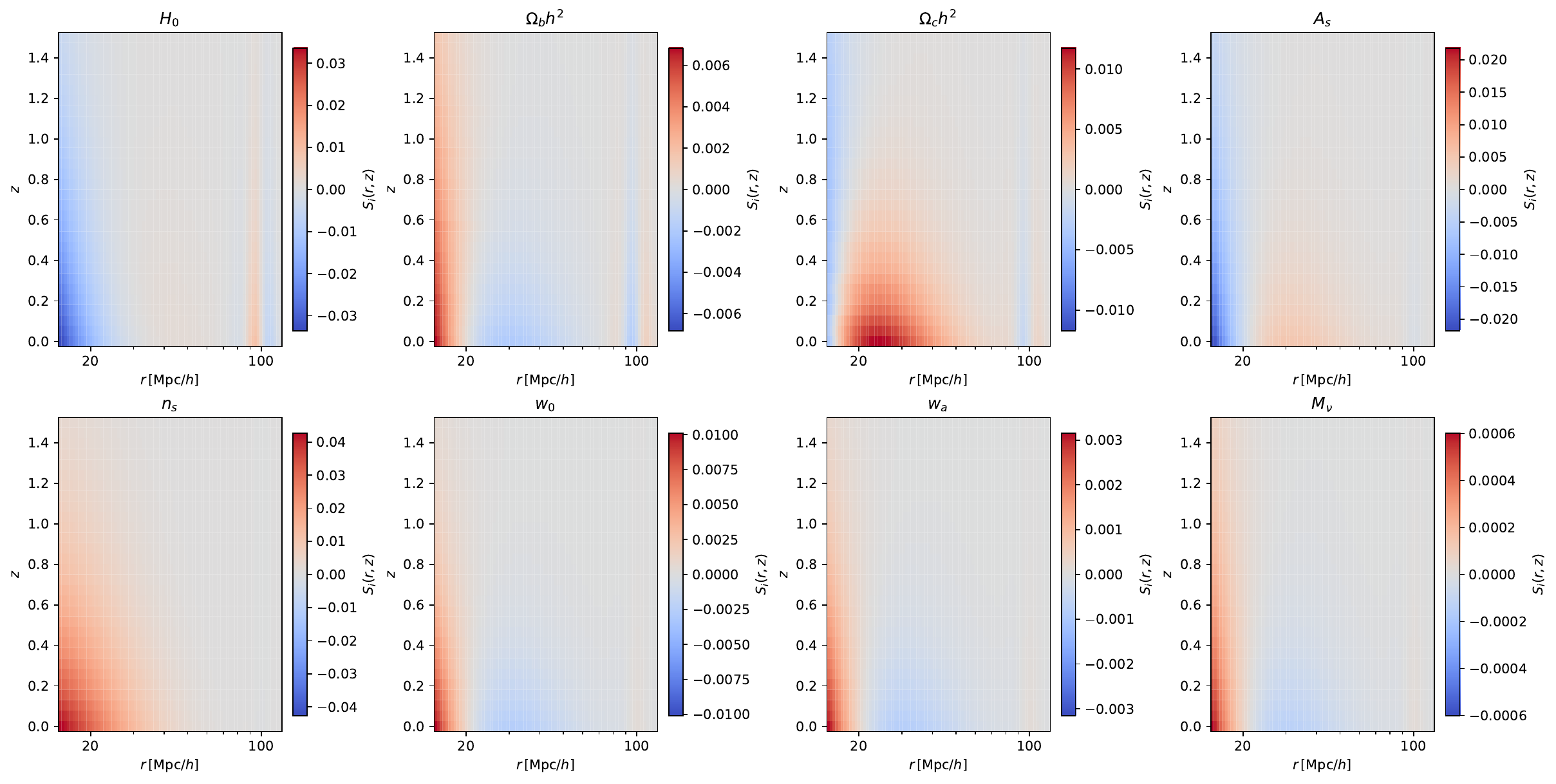}
\caption{Local sensitivities of the nonlinear-to-linear ratio $D_2^{\rm nl}(r,z)/D_2^{\rm lin}(r,z)$. The maps show how each cosmological parameter modifies the relative size of the nonlinear correction to $D_2(r,z)$. The response is again concentrated at small scales and low redshift, with $\Omega_c h^2$, $n_s$ and, more weakly, $m_\nu$ mostly enhancing the ratio, while $A_s$ tends to suppress it.}
\label{fig:ratio_loc_sens}
\end{figure*}

Figure~\ref{fig:ratio_loc_sens} shows the local sensitivities of the ratio $D_2^{\rm nl}(r,z)/D_2^{\rm lin}(r,z)$, with respect to the cosmological parameters. This quantity measures the relative departure of the nonlinear prediction from the linear one. Therefore, its
sensitivity does not describe the total response of $D_2(r,z)$ to a parameter
variation, but rather how that parameter changes the importance of nonlinear
corrections. A positive sensitivity indicates that increasing the corresponding parameter raises the ratio $D_2^{\rm nl}/D_2^{\rm lin}$ towards unity, implying a smaller relative contribution from nonlinear effects. 
The resulting maps show that the parameter dependence of the nonlinear
correction is again concentrated mainly at small scales and low redshift, as
expected from the growth of nonlinear clustering. However, the sign and structure
of the response are not identical to those found for $D_2^{\rm nl}$ itself in
Fig.~\ref{fig:nonlinear_loc_sens}. In particular, $\Omega_c h^2$, $n_s$ and
$m_\nu$ tend to enhance the nonlinear-to-linear ratio in the regions of largest
sensitivity, while $A_s$ mainly suppresses it at lower scales. The dark-energy parameters
show a more scale- and redshift-dependent response, indicating that their effect
on the nonlinear correction cannot be fully described by a simple overall change
in the amplitude of $D_2(r,z)$.

\begin{figure*}
\centering
\includegraphics[width=0.57\textwidth]{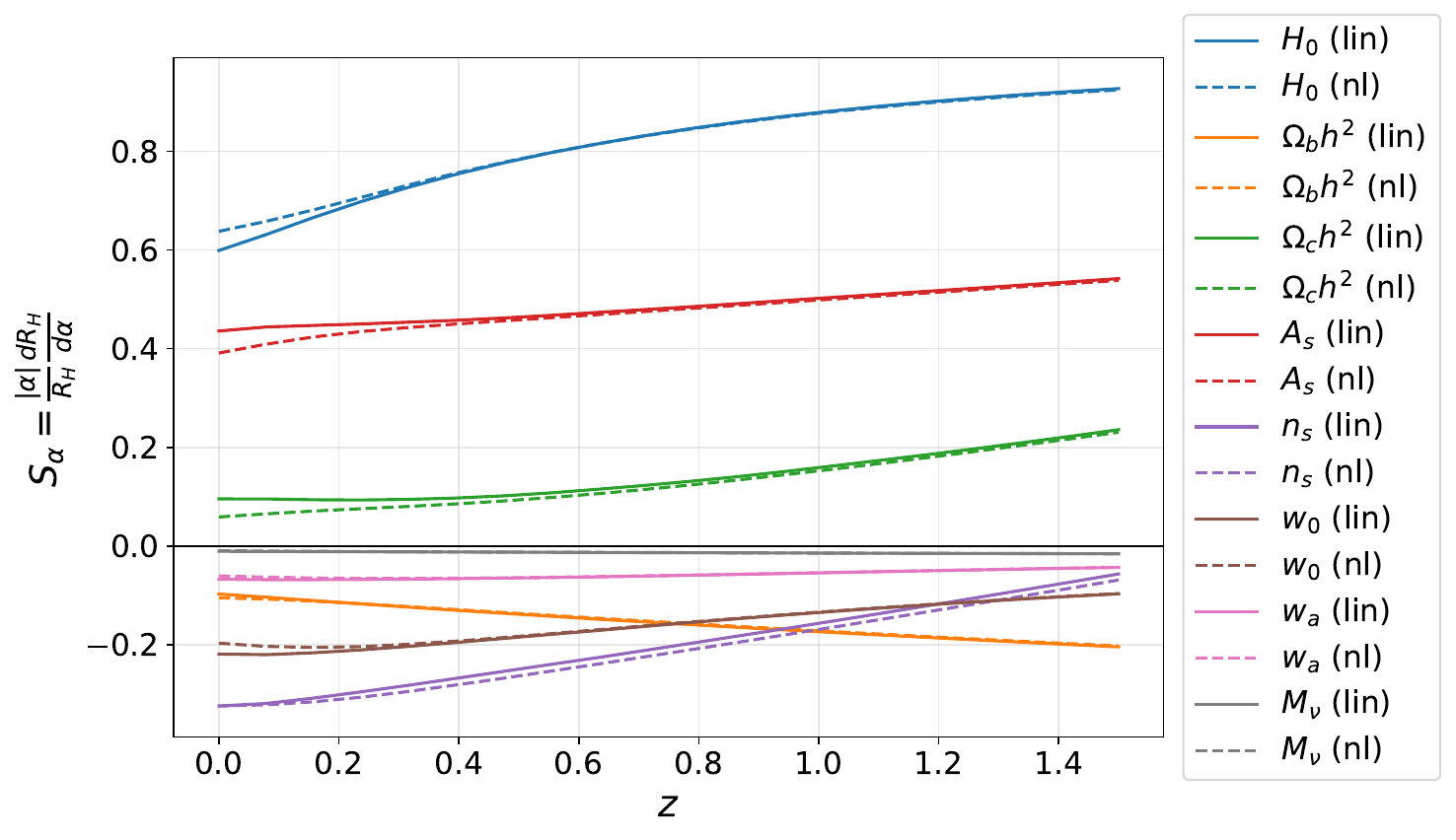}
\includegraphics[width=0.42\textwidth]{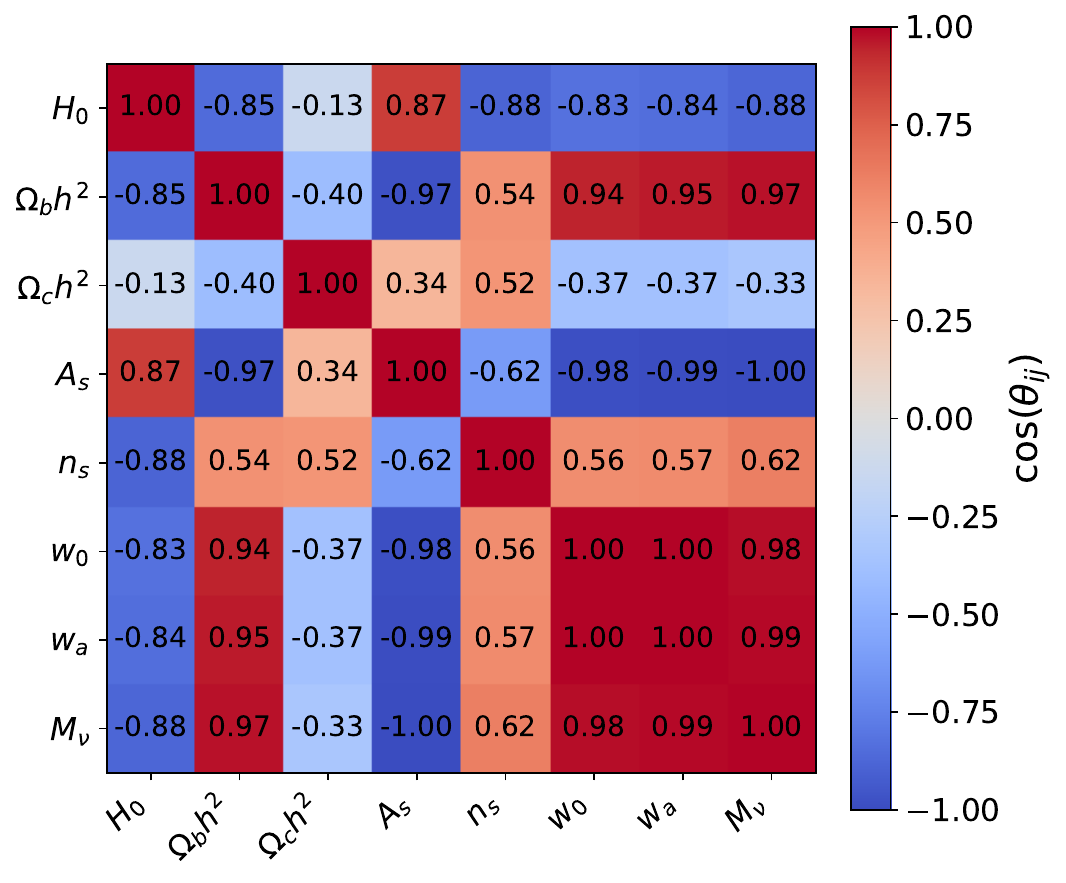}
\caption{{\bf Left.} Local sensitivities of the homogeneity scale $R_{\rm H}(z)$ for the linear and nonlinear predictions. The scale is defined by $D_2(R_{\rm H},z)=2.97$. Increasing $\Omega_c h^2$, $A_s$ and $w_0$ generally shifts the homogeneity scale to larger values, while $\Omega_b h^2$, $n_s$, $w_a$ and $m_\nu$ tend to decrease it. The nonlinear correction only mildly changes these trends. 
{\bf Right.} Angular correlation matrix between the sensitivity vectors of the ratio $D_2^{\rm nl}(r,z)/D_2^{\rm lin} (r,z)$. The matrix shows a strong degeneracy among $\Omega_b h^2$, $A_s$, $w_0$, $w_a$ and $m_\nu$, with correlations or anti-correlations close to unity. By contrast, $\Omega_c h^2$ and $n_s$ display more differentiated responses, with moderate correlations with the remaining parameters. 
}
\label{fig:loc_sens}
\end{figure*}

In addition to the analysis of $D_2(r,z)$ itself, we also study the local
sensitivity of the homogeneity scale $R_{\rm H}(z)$. The results are shown in 
Fig.~\ref{fig:loc_sens} (left panel), where $R_{\rm H}$ is defined through the condition
$D_2(R_{\rm H},z)=2.97$. This provides a compressed description of how changes
in the cosmological parameters affect the transition scale to homogeneity,
rather than the full scale dependence of the correlation dimension. 
The sensitivities of $R_{\rm H}$ show smoother redshift dependences than those
obtained directly for $D_2(r,z)$, since the information is projected onto a
single characteristic scale at each redshift. Increasing $\Omega_c h^2$, $A_s$
and $w_0$ generally shifts $R_{\rm H}$ to larger values, indicating that the
approach to homogeneity occurs at larger scales. By contrast, increasing
$\Omega_b h^2$, $n_s$, $w_a$ and $m_\nu$ tends to reduce $R_{\rm H}$, moving the
transition to homogeneity to smaller scales. The comparison between the linear
and nonlinear curves shows that nonlinear corrections modify the amplitude of
the sensitivity only moderately, without changing the main parameter-dependent
trends.

Finally, we analyze the parameter degeneracies associated with the nonlinear
correction to the correlation dimension. Figure~\ref{fig:loc_sens} (right panel) 
shows the angular correlation matrix constructed from the sensitivity vectors
of the ratio $D_2^{\rm nl}(r,z)/D_2^{\rm lin}(r,z)$. Therefore, this matrix does
not quantify the correlation between the total responses of $D_2(r,z)$, but
rather the similarity between the scale- and redshift-dependent changes induced
by each parameter in the relative nonlinear contribution. 
The matrix reveals that several parameters produce highly degenerate responses
in the nonlinear-to-linear ratio. In particular, $A_s$, $w_0$, $w_a$ and
$m_\nu$, together with $\Omega_b h^2$, form a strongly correlated or
anti-correlated block, with cosine similarities close to $\pm 1$. This indicates
that these parameters modify the nonlinear correction with very similar
patterns in the $(r,z)$ plane, although in some cases with opposite sign. By
contrast, $\Omega_c h^2$ and $n_s$ show more moderate correlations with the
remaining parameters, suggesting a more distinctive scale and redshift
dependence.

\subsection{Global sensitivity analysis of the correlation dimension} 

Our next step is investigating the global sensitivity of the correlation dimension to the cosmological parameters, as a complementary information to the previous local analysis. The goal is to quantify the relative importance of the cosmological parameters using the variance-based Sobol method \cite{SOBOL2001271,JRC40639}. The aim of this Section is to show the impact of the cosmological parameters on the matter power spectrum and its nonlinear corrections, and to compare how such sensitivities are finally imprinted to the correlation dimension in both linear and nonlinear regimes. 
 Since global sensitivity analysis provides a model-agnostic way of identifying which features of an observable are most sensitive to the additional degrees of freedom introduced by an extended model, we will compute and discuss the results for the $\Lambda$CDM and $w_0w_a$CDM cosmologies. 
 
We first start by describing the main aspects of the Sobol methodology. This is a variance-based global sensitivity analysis approach designed to quantify how much of the uncertainty in a model output is attributable to each uncertain input parameter, including non-linear effects and possible parameter interactions. Contrary to the previous local sensitivity analysis, The parameters are considered over their entire prescribed probability distributions, rather than around a single fiducial point. Then, the input parameters are described by independent uniform probability distributions over the corresponding cosmological parameter ranges.
 Given a model $Y=f(\boldsymbol{\theta})$, with inputs $\boldsymbol{\theta}=(\theta_1, \ldots, \theta_d)$ independently and uniformly distributed within the unit hypercube $[0,1]^d$, its variance can be decomposed into contributions from individual parameters and their interactions as 
 $\mathrm{Var}(Y) = \sum_i V_i +\sum_{i<j}V_{ij} + \cdots +V_{1,2,\ldots,d}$,  
 where $V_i={\rm Var}_{\theta_i}(\mathbb{E}_{\theta\sim i}(Y|\theta_i))$ is the first-order partial variance defined as the variance of the conditional expectation of $Y$ given $\theta_i$, $V_{ij}={\rm Var}_{\theta_{ij}}(\mathbb{E}_{\theta_{\sim ij}}[Y|\theta_i,\theta_j])-V_i -V_j$ is the second order partial variance, etc.\footnote{We follow the standard notation $\sim\theta_i$ to indicate the whole set of inputs except $\theta_i$.} The first-order Sobol index of $\theta_i$ on $Y$ measures the fraction of output variance explained by $\theta_i$ alone and is defined as 
 \begin{equation}
     S_i=\frac{V_i}{\mathrm{Var}(Y)}\,.
 \end{equation} 
 
We implement this analysis in our observables of interest following a similar approach  as presented in \cite{Euclid:2018mlb,Euclid:2020rfv}. 
 The general workflow can be summarized in the following steps. 
\begin{enumerate}[label=(\roman*)]
    \item We start by sampling the eight-dimensional cosmological parameter space:
$\boldsymbol{\theta} = \left(A_s,\, n_s,\, \Omega_b,\, \Omega_m,\, h,\, m_\nu,\, w_0,\, w_a\right)$, 
where each parameter is assumed to follow an independent uniform distribution over the fixed intervals of the parameters.
Rather than employing purely random sampling, the cosmological parameter space was sampled using a quasi-random Sobol low-discrepancy sequence \cite{SOBOL196786}, providing an efficient space-filling design in the eight-dimensional parameter space.\footnote{For consistency, the analysis was also tested using Latin Hypercube sampling \cite{Mckay01022000}, finding no significant differences in the final results.}
 For our experiment, we generated $N=1024$ cosmological models, each corresponding to a unique realization of the parameter vector $\boldsymbol{\theta}$.

\item  For each cosmology, the lineal ($P^{\rm lin}(k,z;\boldsymbol{\theta})$) and nonlinear ($P^{\rm nl}(k,z;\boldsymbol{\theta})$) power spectra, as well as the nonlinear correction $B(k,z;\boldsymbol{\theta})\equiv P^{\rm nl}/P^{\rm lin}$, are computed for a given range of scales $k$ and redshifts $z$. From that we then construct the associated linear and nonlinear correlation dimensions ($D_{2}^{\rm lin}(r,z;\boldsymbol{\theta})$ and  $D_2^{\rm lin}(r,z;\boldsymbol{\theta})$, respectively), as well as its nonlinear correction $R(r,z;\boldsymbol{\theta})\equiv D_2^{\rm nl}/D_2^{\rm lin}$. In both cases, the nonlinear-to-linear ratio is expected to isolate the impact of nonlinear gravitational evolution from that of the underlying linear clustering. Each observable is then naturally represented as a function of both scale and redshift for every cosmological realization.

\item For each observable, we construct a matrix 
$M(y,z;\boldsymbol{\theta})\in \mathbb{R}^{N_{\rm cosmo}\times (N_yN_z)}\,$, 
where $y$ denotes either $k$ or $r$, depending on the observable, and $N_y$ and $N_z$ represent the number of points in each grid sample. 
For the case of nonlinear corrections $B$ and $R$, we store their logarithm rather than the quantities themselves. This transformation converts multiplicative variations into additive ones and, for small variations $\Delta\log B\simeq\Delta B/B$, providing a more homogeneous representation of relative variations in the parameter space. We also verified that the overall results are not qualitatively affected by this transformation. 

\item Although each realization contains thousands of values corresponding to different scales and redshifts, much of the information is highly correlated.  To exploit this redundancy, a Principal Component (PC) decomposition is applied to each centered data matrix. Given  $M(y,z;\boldsymbol{\theta})\in \mathbb{R}^{N_{\rm cosmo}\times (N_yN_z)}\,$, the observable is decomposed as 
\begin{equation}
    M(y,z;\boldsymbol{\theta}) \simeq \bar{M}(y,z)+ \sum_{j=1}^{N_{pca}} W_j(\boldsymbol{\theta}){\rm PC}_j(y,z)\,,
\end{equation}
where $W_j$ are the cosmology-dependent PC scores. 
The physical parameters $\boldsymbol{\theta}$ are linearly mapped to normalized variables
$\boldsymbol{\xi}\in[-1,1]^8$ with a uniform distribution, to provide convenient parameterization of the input space.  To choose the number of PCs we can either use a fixed number or impose a variance threshold, although for our proposal this is irrelevant since increasing the number of retained PCs only improves reconstruction by adding higher components, but it should not change PC$_i$ itself. 

\item For each $W_j$, a Polynomial Chaos Expansion (PCE) is constructed w.r.t the normalized parameters $\xi_j$ to fit a Least Angle Regression (LARS) model:
$W_j(\boldsymbol{\xi}) \simeq \widehat{W}_j(\boldsymbol{\xi})=\sum_{k}c_{jk}\Psi_k(\boldsymbol{\xi})\,.$ We use a fixed order, $q=3$, for all PCs. 
Finally, the Sobol indices of each PC score are calculated using their PCE surrogate model.  The first-order index of the parameter $\theta_i$ in $W_j$ is then computed as 
$S_i=\frac{{\rm Var}_{\xi_i}[\mathbb{E}(\widehat{W}_j|\xi_i)]}{{\rm Var}(\widehat{W}_j)}\,.$
\end{enumerate}
All the results shown below were obtained using matter power spectra and their corresponding nonlinear corrections from EE2 within the scales $k\in[9\times 10^{-3}, 9.4]$\,$h$/Mpc and redshifts within the range $[0,3]$, and a parameter space within the ranges detailed in Table~\ref{tab:EE2ranges}, for which the emulator is well-defined. As a self consistency of our analysis, we also have computed the same global sensitivity analysis using HMCode2020, finding no significant differences.

\begin{table*}
\centering
\setlength{\tabcolsep}{9pt}
\begin{tabular}{lccccccccc}
\hline\hline
        & $A_s$ & $n_s$  & $\Omega_b$ & $\Omega_m$  & $h$ & $m_\nu$ & $w_0$ & $w_a$ \\ \hline
Minimum & $1.7\times 10^{-9}$ & $0.92$ & $0.04$ & $0.24$ & $0.61$ & $0$\,eV & $-1.3$ &  $-0.7$  \\
Maximum  & $2.5\times 10^{-9}$ & $1$ & $0.06$ & $0.40$ & $0.73$  & $0.15$\,eV & $-0.7$ & $0.5$  \\
\hline\hline
\end{tabular}
\caption{
Parameter ranges used for the global sensitivity analysis. Matter power spectra and their corresponding nonlinear corrections were computed using EE2 within the scales $k \in [9 \times 10^{-3}, 9.4]\, h/$Mpc and redshifts within the range $[0, 3]$.
}
\label{tab:EE2ranges}
\end{table*}

\begin{figure*}
\centering
\includegraphics[width=0.99\textwidth]{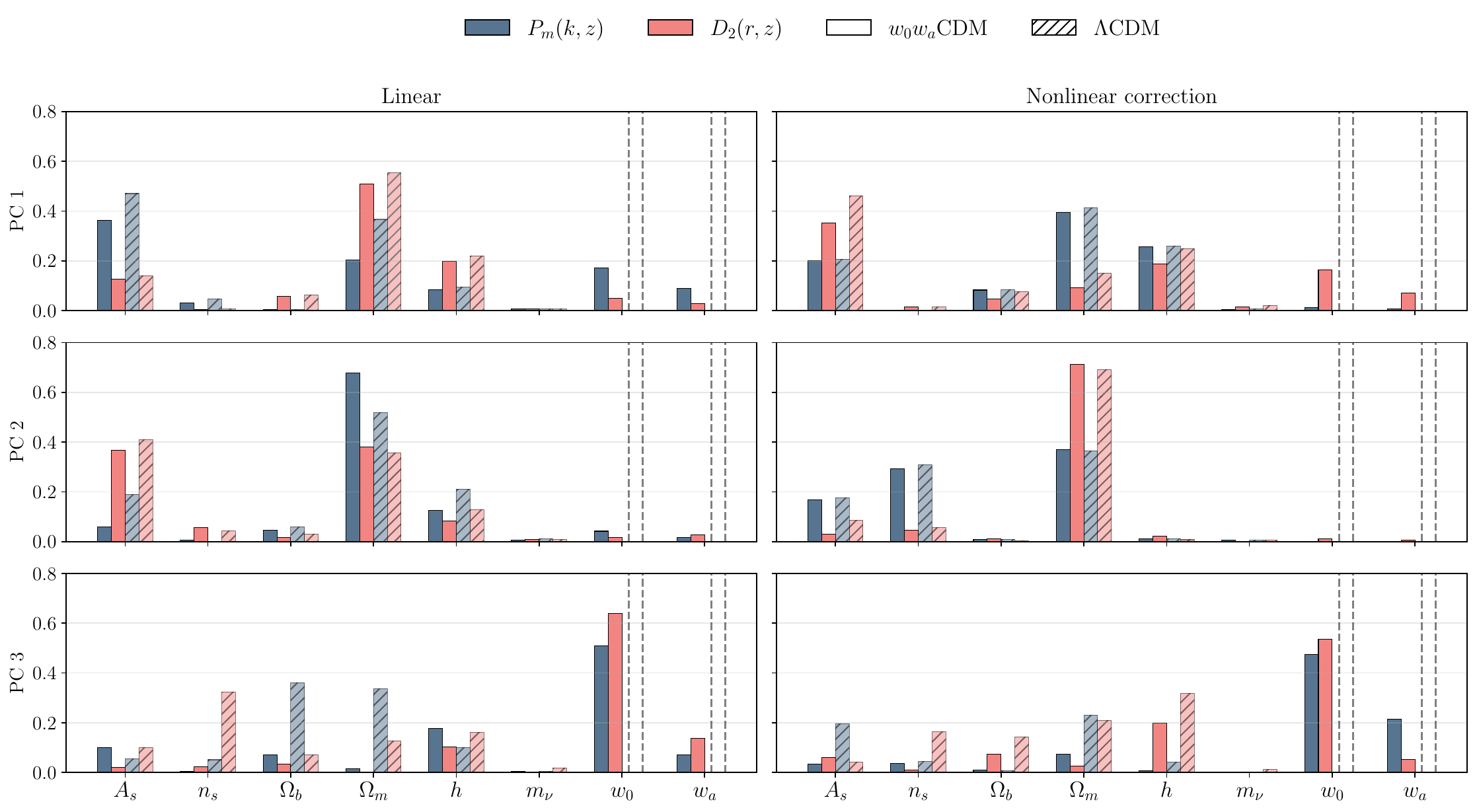}
\caption{Global sensitivity using Sobol analysis for the power spectra (blue) and correlation dimensions (coral) for the $w_0w_a$CDM model (fulfill bars) in the corresponding scales within the ranges $k\in [8.73\times 10^{-3}, 9.41]\,h$/Mpc and $r\in[10,200]\,$Mpc/$h$. The linear and nonlinear cases are display in the first and second columns, respectively, while their corresponding nonlinear corrections ($B(k,z)$ and $R(r,z)$) are shown in the third column. In addition, the sensitivities obtained for the same observables but computed for the $\Lambda$CDM model (striped bars) are also shown, for which $w_0=-1$ and $w_a=0$ (these fixed parameters are indicated with vertical dashed lines).}
\label{fig:SobolPkvsD2}
\end{figure*}

We first compare in Fig.~\ref{fig:SobolPkvsD2} and Table~\ref{tab:sensitivities} the sensitivity on the cosmological parameters of the linear and nonlinear matter power spectra and the correlation dimensions, as well as their associated nonlinear corrections $B(k,z)$ and $R(r,z)$, for the corresponding scales within the ranges $k\in [8.73\times 10^{-3}, 9.41]$\,$h$/Mpc and $r\in[10,200]\,$Mpc/$h$. Although this comparison is not straightforward, since the computation of the correlation dimension involves a Fourier-space transformation of the matter power spectrum, this preliminary analysis can provide insight into how the sensitivity to the cosmological parameters is transferred from one observable to the other, even in the linear regime. We show here the first three PCs, which together account for more than 0.99 
of the explained cumulative variance in all the cases analyzed. The results are displayed for both the $\LCDM$ and $w_0w_a$CDM models. 
 Regarding the matter power spectra, the nonlinear correction pattern for the $w_0w_a$CDM model, $B(k,z;\boldsymbol{\theta})$, is in good agreement with that reported in \cite{Euclid:2020rfv}, supporting the consistency of our method. 
Compared to the $\LCDM$ model, the sensitivity of $B(k,z;\boldsymbol{\theta})$ on the cosmological parameters is almost the same for both models for PC$_1$ ($\Omega_m>h>A_s>\ldots$) and PC$_2$ ($\Omega_m>h>A_s>\ldots$), while the additional cosmological parameters $w_0$ and $w_a$ become quite relevant at the level of PC$_3$ ($w_0>w_a>\Omega_m>\ldots$) as a distinctive feature of the $w_0w_a$CDM model. On the other hand, the sensitivity of the correlation dimension in the linear regime is dominated by the parameters $\Omega_m$, $h$ and $A_s$ in the two cosmological scenarios. Moreover, in this case the additional parameter $w_0$ also becomes relevant at the level of the PC$_1$ for nonlinear correction $R(k,z;\boldsymbol{\theta})$. In general, these comparisons could provide a way to assess whether the nonlinear scales in the correlation dimension contain additional information capable of distinguishing extended cosmological models from $\Lambda$CDM. We also verify that second-order  interactions between the parameters are negligible. 

\begin{figure*}
\centering
\includegraphics[width=0.99\textwidth]{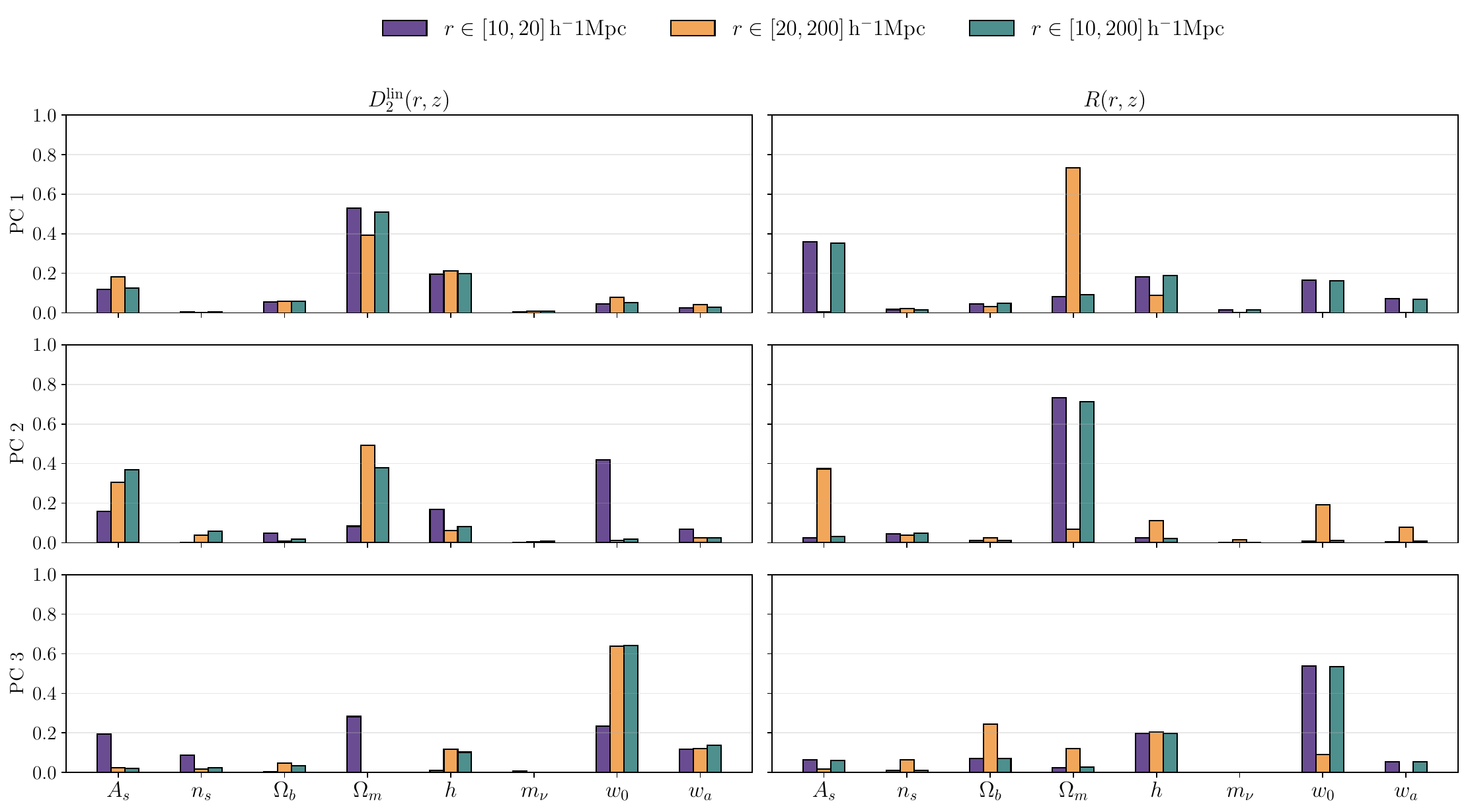}
\caption{Global sensitivity using Sobol analysis for $D_2^{\rm lin}(r,z;\boldsymbol{\theta})$ and the non-linear correction $R(r,z;\boldsymbol{\theta})$, and for scales within the ranges $r\in[10,20]\,$h$^{-1}$Mpc, $r\in[20,200]\,$h$^{-1}$Mpc and $r\in[10,200]\,$h$^{-1}$Mpc. }
\label{fig:SobolEE2}
\end{figure*}

An interesting analysis of the global sensitivities is also the possibility of identifying the scales where the additional degrees of freedom of extended cosmological models (through the parameters $w_0$ and $w_a$, in our case) leave the strongest imprint on the observable. In this direction, and to better distinguish the contributions tin the linear and nonlinear regimes, we show in Fig.~\ref{fig:SobolEE2} the global sensitivity of $D_2(r,z)$ and  $R(r,z)$ for the scales within the ranges $r\in[10,20]\,$Mpc/$h$, $r\in[20,200]\,$Mpc/$h$ and $r\in[10,200]\,$Mpc/$h$ separately. As expected, the sensitivities for $r\in[10,20]\,$Mpc/$h$ and $r\in[10,200]\,$Mpc/$h$ exhibit the same pattern, indicating that the dominant contributions to these observables arise from the smallest scales, $r\in[10,20]\,$Mpc/$h$. Furthermore, the nonlinear correction appears to be more sensitive to the range of $r$, with the relative importance of the cosmological parameters changing as the range of scales is extended.

\begin{table*}
\centering
\begin{tabular}{@{\extracolsep{15pt}}clll}
\hline \hline
  $\Lambda$CDM &  Sensitivity PC$_1$ & Sensitivity PC$_2$ &Sensitivity PC$_3$ \\
\hline
$P_{\rm lin}(k,z;\boldsymbol{\theta})$  &$A_s>\Omega_m>h>\ldots$ & $\Omega_m>h>A_s>\ldots$ & $\Omega_b>\Omega_m>h>\ldots$ \\
$B(k,z;\boldsymbol{\theta})$ & $\Omega_m>h>A_s>\ldots$ & $\Omega_m>n_s>A_s>\ldots$ & $\Omega_m>A_s>n_s>\ldots$ \\
$D_{2}^{\rm lin}(r,z;\boldsymbol{\theta})$  &$\Omega_m>h>A_s>\ldots$ & $A_s>\Omega_m>h>\ldots$ & $n_s>h>\Omega_m>\ldots$ \\
$R(r,z;\boldsymbol{\theta})$  &$A_s>h>\Omega_m>\ldots$ & $\Omega_m>A_s>n_s>\ldots$  & $h>\Omega_m>n_s>\ldots$  \\
\hline\hline
  $w_0w_a$CDM  &Sensitivity PC$_1$ & Sensitivity PC$_2$ &Sensitivity PC$_3$ \\
\hline
$P_{\rm lin}(k,z;\boldsymbol{\theta})$  & $A_s>\Omega_m>w_0>\ldots$ & $\Omega_m>h>A_s>\ldots$ & $w_0>h>A_s>\ldots$\\
$B(k,z;\boldsymbol{\theta})$ & $\Omega_m>h>A_s>\ldots$ & $\Omega_m>n_s>A_s>\ldots$ & $w_0>w_a>\Omega_m>\ldots$ \\
$D_{2}^{\rm lin}(r,z;\boldsymbol{\theta})$ & $\Omega_m>h>A_s>\ldots$ & $\Omega_m>A_s>h>\ldots$ & $w_0>w_a>h>\ldots$ \\
$R(r,z;\boldsymbol{\theta})$   & $A_s>h>w_0>\ldots$ & $\Omega_m>n_s>A_s>\ldots$ & $w_0>h>\Omega_b>\ldots$\\
\hline\hline
\end{tabular}
\caption{Summary of the sensitivity information displayed in Fig.~\ref{fig:SobolEE2} for both $\LCDM$ and $w_0w_a$CDM models. For each observable, the first three parameters are shown in order of their Sobol sensitivity importance. The nonlinear correction of the matter power spectra, $B(k,z;\boldsymbol{\theta})$, are in agreement with those the reported in \cite{Euclid:2020rfv}.
}
\label{tab:sensitivities}
\end{table*}

\section{Discussion}\label{discussion}

In this work, we have investigated the effects of non-linear structure formation on the correlation dimension $D_2(r,z)$, and its cosmological dependence within the $\Lambda$CDM and $w_0w_a$CDM frameworks. We have developed and calibrated an analytical approximation connecting $D_2$ to the matter power spectrum and combined local derivative-based and global variance-based sensitivity analyzes to identify the parameters that govern its scale and redshift dependence.

For the cosmologies and scales examined, the non-linear enhancement of small-scale power leads to a lower correlation dimension than the linear prediction. Non-linear evolution therefore strengthens the departure from homogeneous scaling on these scales, as quantified by the increased difference between $D_2$ and the homogeneous value of three. This behavior remains compatible with the approach to large-scale homogeneity, as both predictions approach $D_2= 3$ at a large counting radius. Non-linear departures become increasingly pronounced at radii of order $20\,$Mpc/$h$ and below, particularly at low redshift. This scale characterizes the cases examined and should not be interpreted as a universal threshold independent of cosmology, redshift, or the required modeling accuracy.

The calibrated cut-off approximation provides an analytical interpretation of these results by expressing the departure from homogeneity as the dimensionless power at an effective Fourier cut-off, $k_{\rm cut}=\alpha/r$, divided by unity plus the accumulated power below that cut-off. The effect of non-linear evolution consequently depends on the relative changes in these two contributions, rather than on the enhancement of power alone. With $\alpha\simeq2.39$, the approximation closely reproduces the exact correlation dimension, with relative errors below one per cent in the representative linear and non-linear examples examined. This accuracy refers to $D_2$ itself and does not imply the same relative accuracy for the smaller departure $3-D_2$ or for the non-linear correction. By replacing the oscillatory integral expressions with a cumulative integral and an evaluation of the spectrum at the effective cut-off, the approximation offers a simpler computational prescription while retaining a transparent connection to the underlying clustering.

Sensitivity analyzes highlight the importance of matter density and primordial amplitude in determining the behavior of $D_2$. The local responses are generally strongest at small radii and low redshift and weaken towards the homogeneous regime. In the global analysis, $\Omega_m$ dominates the leading principal component of the linear correlation dimension, whereas $A_s$ is the largest contributor to the leading component of the non-linear-to-linear ratio. The Hubble parameter also contributes appreciably. These findings identify matter content and primordial amplitude as key influences. The resulting hierarchy also depends on the observable, the principal component, and the parameter ranges and distributions adopted in the global analysis.

The dark-energy parameters $w_0$ and $w_a$ produce scale- and redshift-dependent changes in the relative non-linear contribution that cannot generally be represented by a uniform rescaling. Similarities between their local response patterns and those of other parameters indicate potential degeneracies, although angular similarities alone do not establish posterior correlations. Within the explored domain, the comparatively small global contributions of $n_s$, $\Omega_b$ and, especially, $m_\nu$ motivate investigating reduced parameter spaces for future fits. The weak sensitivity to neutrino mass makes $m_\nu$ a particularly relevant candidate for external constraints or a fixed value, provided that doing so does not appreciably shift the inferred parameters. More caution is required for $n_s$, whose modest global contribution coexists with an appreciable local response. These results therefore guide which simplifications should be tested, rather than establishing that the corresponding parameters can already be neglected.

The present analysis is limited to the theoretical matter correlation dimension and does not include the full mapping to galaxy-survey measurements, including Alcock–Paczyński distortions, linear galaxy bias, coherent redshift-space distortions and small-scale velocity dispersion through the Fingers-of-God effect \cite{Ntelis:2017nrj, Ntelis:2018ctq}. It isolates the cosmological dependence associated with the underlying matter clustering, which is a necessary component of the full observational prediction. Future work will extend the sensitivity analyzes to the tracer observable with these effects and their associated nuisance parameters, assessing whether the parameter hierarchy and proposed simplifications remain valid. Incorporating realistic survey covariances will then allow these results to guide the choice of fitting scales, the treatment of weakly constrained parameters, and the use of external information, providing a basis for future cosmological fits to measured $D_2(r,z)$.

\FloatBarrier
\begin{acknowledgments}
FATP acknowledges the support of the grant PID2024-158938NB-I00 funded by MICIU/AEI/ 10.13039/501100011033 and by “ERDF A way of making Europe”, and the grant from Project SA097P24 funded by Junta de Castilla y Le\'on.
\end{acknowledgments}

\bibliography{bib}

\begin{thebibliography}{29}%
\makeatletter
\providecommand \@ifxundefined [1]{%
 \@ifx{#1\undefined}
}%
\providecommand \@ifnum [1]{%
 \ifnum #1\expandafter \@firstoftwo
 \else \expandafter \@secondoftwo
 \fi
}%
\providecommand \@ifx [1]{%
 \ifx #1\expandafter \@firstoftwo
 \else \expandafter \@secondoftwo
 \fi
}%
\providecommand \natexlab [1]{#1}%
\providecommand \enquote  [1]{``#1''}%
\providecommand \bibnamefont  [1]{#1}%
\providecommand \bibfnamefont [1]{#1}%
\providecommand \citenamefont [1]{#1}%
\providecommand \href@noop [0]{\@secondoftwo}%
\providecommand \href [0]{\begingroup \@sanitize@url \@href}%
\providecommand \@href[1]{\@@startlink{#1}\@@href}%
\providecommand \@@href[1]{\endgroup#1\@@endlink}%
\providecommand \@sanitize@url [0]{\catcode `\\12\catcode `\$12\catcode
  `\&12\catcode `\#12\catcode `\^12\catcode `\_12\catcode `\%12\relax}%
\providecommand \@@startlink[1]{}%
\providecommand \@@endlink[0]{}%
\providecommand \url  [0]{\begingroup\@sanitize@url \@url }%
\providecommand \@url [1]{\endgroup\@href {#1}{\urlprefix }}%
\providecommand \urlprefix  [0]{URL }%
\providecommand \Eprint [0]{\href }%
\providecommand \doibase [0]{http://dx.doi.org/}%
\providecommand \selectlanguage [0]{\@gobble}%
\providecommand \bibinfo  [0]{\@secondoftwo}%
\providecommand \bibfield  [0]{\@secondoftwo}%
\providecommand \translation [1]{[#1]}%
\providecommand \BibitemOpen [0]{}%
\providecommand \bibitemStop [0]{}%
\providecommand \bibitemNoStop [0]{.\EOS\space}%
\providecommand \EOS [0]{\spacefactor3000\relax}%
\providecommand \BibitemShut  [1]{\csname bibitem#1\endcsname}%
\let\auto@bib@innerbib\@empty
\bibitem [{\citenamefont {Hentschel}\ and\ \citenamefont
  {Procaccia}(1983)}]{Hentschel:1983zhc}%
  \BibitemOpen
  \bibfield  {author} {\bibinfo {author} {\bibfnamefont {H.~G.~E.}\
  \bibnamefont {Hentschel}}\ and\ \bibinfo {author} {\bibfnamefont
  {I.}~\bibnamefont {Procaccia}},\ }\href {\doibase
  10.1016/0167-2789(83)90235-X} {\bibfield  {journal} {\bibinfo  {journal}
  {Physica}\ }\textbf {\bibinfo {volume} {8}},\ \bibinfo {pages} {435}
  (\bibinfo {year} {1983})}\BibitemShut {NoStop}%
\bibitem [{\citenamefont {Borgani}(1995)}]{Borgani:1994uy}%
  \BibitemOpen
  \bibfield  {author} {\bibinfo {author} {\bibfnamefont {S.}~\bibnamefont
  {Borgani}},\ }\href {\doibase 10.1016/0370-1573(94)00073-C} {\bibfield
  {journal} {\bibinfo  {journal} {Phys. Rept.}\ }\textbf {\bibinfo {volume}
  {251}},\ \bibinfo {pages} {1} (\bibinfo {year} {1995})},\ \Eprint
  {http://arxiv.org/abs/astro-ph/9404054} {arXiv:astro-ph/9404054} \BibitemShut
  {NoStop}%
\bibitem [{\citenamefont {Gaite}(2019)}]{Gaite:2018kbu}%
  \BibitemOpen
  \bibfield  {author} {\bibinfo {author} {\bibfnamefont {J.}~\bibnamefont
  {Gaite}},\ }\href {\doibase 10.1155/2019/6587138} {\bibfield  {journal}
  {\bibinfo  {journal} {Adv. Astron.}\ }\textbf {\bibinfo {volume} {2019}},\
  \bibinfo {pages} {6587138} (\bibinfo {year} {2019})},\ \Eprint
  {http://arxiv.org/abs/1810.02311} {arXiv:1810.02311 [astro-ph.CO]}
  \BibitemShut {NoStop}%
\bibitem [{\citenamefont {Yadav}\ \emph {et~al.}(2010)\citenamefont {Yadav},
  \citenamefont {Bagla},\ and\ \citenamefont {Khandai}}]{Yadav:2010cc}%
  \BibitemOpen
  \bibfield  {author} {\bibinfo {author} {\bibfnamefont {J.~K.}\ \bibnamefont
  {Yadav}}, \bibinfo {author} {\bibfnamefont {J.~S.}\ \bibnamefont {Bagla}}, \
  and\ \bibinfo {author} {\bibfnamefont {N.}~\bibnamefont {Khandai}},\ }\href
  {\doibase 10.1111/j.1365-2966.2010.16612.x} {\bibfield  {journal} {\bibinfo
  {journal} {Mon. Not. Roy. Astron. Soc.}\ }\textbf {\bibinfo {volume} {405}},\
  \bibinfo {pages} {2009} (\bibinfo {year} {2010})},\ \Eprint
  {http://arxiv.org/abs/1001.0617} {arXiv:1001.0617 [astro-ph.CO]} \BibitemShut
  {NoStop}%
\bibitem [{\citenamefont {Einasto}\ \emph {et~al.}(2020)\citenamefont
  {Einasto}, \citenamefont {H{\"u}tsi}, \citenamefont {Kuutma},\ and\
  \citenamefont {Einasto}}]{Einasto:2020jrp}%
  \BibitemOpen
  \bibfield  {author} {\bibinfo {author} {\bibfnamefont {J.}~\bibnamefont
  {Einasto}}, \bibinfo {author} {\bibfnamefont {G.}~\bibnamefont {H{\"u}tsi}},
  \bibinfo {author} {\bibfnamefont {T.}~\bibnamefont {Kuutma}}, \ and\ \bibinfo
  {author} {\bibfnamefont {M.}~\bibnamefont {Einasto}},\ }\href {\doibase
  10.1051/0004-6361/202037683} {\bibfield  {journal} {\bibinfo  {journal}
  {Astron. Astrophys.}\ }\textbf {\bibinfo {volume} {640}},\ \bibinfo {pages}
  {A47} (\bibinfo {year} {2020})},\ \Eprint {http://arxiv.org/abs/2002.02813}
  {arXiv:2002.02813 [astro-ph.CO]} \BibitemShut {NoStop}%
\bibitem [{\citenamefont {{Hogg}}\ \emph {et~al.}(2005)\citenamefont {{Hogg}}
  \emph {et~al.}}]{Hogg:2004vw}%
  \BibitemOpen
  \bibfield  {author} {\bibinfo {author} {\bibfnamefont {D.~W.}\ \bibnamefont
  {{Hogg}}} \emph {et~al.},\ }\href {\doibase 10.1086/429084} {\bibfield
  {journal} {\bibinfo  {journal} {\apj}\ }\textbf {\bibinfo {volume} {624}},\
  \bibinfo {pages} {54} (\bibinfo {year} {2005})},\ \Eprint
  {http://arxiv.org/abs/astro-ph/0411197} {arXiv:astro-ph/0411197 [astro-ph]}
  \BibitemShut {NoStop}%
\bibitem [{\citenamefont {Sarkar}\ \emph {et~al.}(2009)\citenamefont {Sarkar},
  \citenamefont {Yadav}, \citenamefont {Pandey},\ and\ \citenamefont
  {Bharadwaj}}]{Sarkar_2009}%
  \BibitemOpen
  \bibfield  {author} {\bibinfo {author} {\bibfnamefont {P.}~\bibnamefont
  {Sarkar}}, \bibinfo {author} {\bibfnamefont {J.}~\bibnamefont {Yadav}},
  \bibinfo {author} {\bibfnamefont {B.}~\bibnamefont {Pandey}}, \ and\ \bibinfo
  {author} {\bibfnamefont {S.}~\bibnamefont {Bharadwaj}},\ }\href {\doibase
  10.1111/j.1745-3933.2009.00738.x} {\bibfield  {journal} {\bibinfo  {journal}
  {Monthly Notices of the Royal Astronomical Society: Letters}\ }\textbf
  {\bibinfo {volume} {399}},\ \bibinfo {pages} {L128–L131} (\bibinfo {year}
  {2009})}\BibitemShut {NoStop}%
\bibitem [{\citenamefont {{Scrimgeour}}\ \emph {et~al.}(2012)\citenamefont
  {{Scrimgeour}} \emph {et~al.}}]{Scrimgeour:2012wt}%
  \BibitemOpen
  \bibfield  {author} {\bibinfo {author} {\bibfnamefont {M.~I.}\ \bibnamefont
  {{Scrimgeour}}} \emph {et~al.},\ }\href {\doibase
  10.1111/j.1365-2966.2012.21402.x} {\bibfield  {journal} {\bibinfo  {journal}
  {\mnras}\ }\textbf {\bibinfo {volume} {425}},\ \bibinfo {pages} {116}
  (\bibinfo {year} {2012})},\ \Eprint {http://arxiv.org/abs/1205.6812}
  {arXiv:1205.6812 [astro-ph.CO]} \BibitemShut {NoStop}%
\bibitem [{\citenamefont {Goyal}\ \emph {et~al.}(2024)\citenamefont {Goyal},
  \citenamefont {Malik}, \citenamefont {Yadav},\ and\ \citenamefont
  {Seshadri}}]{Goyal:2024ctd}%
  \BibitemOpen
  \bibfield  {author} {\bibinfo {author} {\bibfnamefont {P.}~\bibnamefont
  {Goyal}}, \bibinfo {author} {\bibfnamefont {S.}~\bibnamefont {Malik}},
  \bibinfo {author} {\bibfnamefont {J.~k.}\ \bibnamefont {Yadav}}, \ and\
  \bibinfo {author} {\bibfnamefont {T.~R.}\ \bibnamefont {Seshadri}},\ }\href
  {\doibase 10.1093/mnras/stae1041} {\bibfield  {journal} {\bibinfo  {journal}
  {Mon. Not. Roy. Astron. Soc.}\ }\textbf {\bibinfo {volume} {530}},\ \bibinfo
  {pages} {2866} (\bibinfo {year} {2024})},\ \Eprint
  {http://arxiv.org/abs/2404.09197} {arXiv:2404.09197 [astro-ph.CO]}
  \BibitemShut {NoStop}%
\bibitem [{\citenamefont {Chevallier}\ and\ \citenamefont
  {Polarski}(2001)}]{Chevallier:2000qy}%
  \BibitemOpen
  \bibfield  {author} {\bibinfo {author} {\bibfnamefont {M.}~\bibnamefont
  {Chevallier}}\ and\ \bibinfo {author} {\bibfnamefont {D.}~\bibnamefont
  {Polarski}},\ }\href {\doibase 10.1142/S0218271801000822} {\bibfield
  {journal} {\bibinfo  {journal} {Int. J. Mod. Phys. D}\ }\textbf {\bibinfo
  {volume} {10}},\ \bibinfo {pages} {213} (\bibinfo {year} {2001})},\ \Eprint
  {http://arxiv.org/abs/gr-qc/0009008} {arXiv:gr-qc/0009008} \BibitemShut
  {NoStop}%
\bibitem [{\citenamefont {Linder}(2003)}]{Linder:2002et}%
  \BibitemOpen
  \bibfield  {author} {\bibinfo {author} {\bibfnamefont {E.~V.}\ \bibnamefont
  {Linder}},\ }\href {\doibase 10.1103/PhysRevLett.90.091301} {\bibfield
  {journal} {\bibinfo  {journal} {Phys. Rev. Lett.}\ }\textbf {\bibinfo
  {volume} {90}},\ \bibinfo {pages} {091301} (\bibinfo {year} {2003})},\
  \Eprint {http://arxiv.org/abs/astro-ph/0208512} {arXiv:astro-ph/0208512}
  \BibitemShut {NoStop}%
\bibitem [{\citenamefont {Abdul~Karim}\ \emph {et~al.}(2025)\citenamefont
  {Abdul~Karim} \emph {et~al.}}]{DESIdr2}%
  \BibitemOpen
  \bibfield  {author} {\bibinfo {author} {\bibfnamefont {M.}~\bibnamefont
  {Abdul~Karim}} \emph {et~al.} (\bibinfo {collaboration} {DESI
  Collaboration}),\ }\href {\doibase 10.1103/tr6y-kpc6} {\bibfield  {journal}
  {\bibinfo  {journal} {Phys. Rev. D}\ }\textbf {\bibinfo {volume} {112}},\
  \bibinfo {pages} {083515} (\bibinfo {year} {2025})}\BibitemShut {NoStop}%
\bibitem [{\citenamefont {Mead}\ \emph {et~al.}(2015)\citenamefont {Mead},
  \citenamefont {Peacock}, \citenamefont {Heymans}, \citenamefont {Joudaki},\
  and\ \citenamefont {Heavens}}]{Mead:2015yca}%
  \BibitemOpen
  \bibfield  {author} {\bibinfo {author} {\bibfnamefont {A.}~\bibnamefont
  {Mead}}, \bibinfo {author} {\bibfnamefont {J.}~\bibnamefont {Peacock}},
  \bibinfo {author} {\bibfnamefont {C.}~\bibnamefont {Heymans}}, \bibinfo
  {author} {\bibfnamefont {S.}~\bibnamefont {Joudaki}}, \ and\ \bibinfo
  {author} {\bibfnamefont {A.}~\bibnamefont {Heavens}},\ }\href {\doibase
  10.1093/mnras/stv2036} {\bibfield  {journal} {\bibinfo  {journal} {Mon. Not.
  Roy. Astron. Soc.}\ }\textbf {\bibinfo {volume} {454}},\ \bibinfo {pages}
  {1958} (\bibinfo {year} {2015})},\ \Eprint {http://arxiv.org/abs/1505.07833}
  {arXiv:1505.07833 [astro-ph.CO]} \BibitemShut {NoStop}%
\bibitem [{\citenamefont {Mead}\ \emph {et~al.}(2021)\citenamefont {Mead},
  \citenamefont {Brieden}, \citenamefont {Tr{\"o}ster},\ and\ \citenamefont
  {Heymans}}]{Mead:2020vgs}%
  \BibitemOpen
  \bibfield  {author} {\bibinfo {author} {\bibfnamefont {A.}~\bibnamefont
  {Mead}}, \bibinfo {author} {\bibfnamefont {S.}~\bibnamefont {Brieden}},
  \bibinfo {author} {\bibfnamefont {T.}~\bibnamefont {Tr{\"o}ster}}, \ and\
  \bibinfo {author} {\bibfnamefont {C.}~\bibnamefont {Heymans}},\ }\href
  {\doibase 10.1093/mnras/stab082} {\bibfield  {journal} {\bibinfo  {journal}
  {Mon. Not. Roy. Astron. Soc.}\ }\textbf {\bibinfo {volume} {502}},\ \bibinfo
  {pages} {1401} (\bibinfo {year} {2021})},\ \Eprint
  {http://arxiv.org/abs/2009.01858} {arXiv:2009.01858 [astro-ph.CO]}
  \BibitemShut {NoStop}%
\bibitem [{\citenamefont {Knabenhans}\ \emph {et~al.}(2019)\citenamefont
  {Knabenhans} \emph {et~al.}}]{Euclid:2018mlb}%
  \BibitemOpen
  \bibfield  {author} {\bibinfo {author} {\bibfnamefont {M.}~\bibnamefont
  {Knabenhans}} \emph {et~al.} (\bibinfo {collaboration} {Euclid}),\ }\href
  {\doibase 10.1093/mnras/stz197} {\bibfield  {journal} {\bibinfo  {journal}
  {Mon. Not. Roy. Astron. Soc.}\ }\textbf {\bibinfo {volume} {484}},\ \bibinfo
  {pages} {5509} (\bibinfo {year} {2019})},\ \Eprint
  {http://arxiv.org/abs/1809.04695} {arXiv:1809.04695 [astro-ph.CO]}
  \BibitemShut {NoStop}%
\bibitem [{\citenamefont {Knabenhans}\ \emph {et~al.}(2021)\citenamefont
  {Knabenhans} \emph {et~al.}}]{Euclid:2020rfv}%
  \BibitemOpen
  \bibfield  {author} {\bibinfo {author} {\bibfnamefont {M.}~\bibnamefont
  {Knabenhans}} \emph {et~al.} (\bibinfo {collaboration} {Euclid}),\ }\href
  {\doibase 10.1093/mnras/stab1366} {\bibfield  {journal} {\bibinfo  {journal}
  {Mon. Not. Roy. Astron. Soc.}\ }\textbf {\bibinfo {volume} {505}},\ \bibinfo
  {pages} {2840} (\bibinfo {year} {2021})},\ \Eprint
  {http://arxiv.org/abs/2010.11288} {arXiv:2010.11288 [astro-ph.CO]}
  \BibitemShut {NoStop}%
\bibitem [{\citenamefont {Dvali}\ \emph {et~al.}(2000)\citenamefont {Dvali},
  \citenamefont {Gabadadze},\ and\ \citenamefont {Porrati}}]{Dvali:2000hr}%
  \BibitemOpen
  \bibfield  {author} {\bibinfo {author} {\bibfnamefont {G.~R.}\ \bibnamefont
  {Dvali}}, \bibinfo {author} {\bibfnamefont {G.}~\bibnamefont {Gabadadze}}, \
  and\ \bibinfo {author} {\bibfnamefont {M.}~\bibnamefont {Porrati}},\ }\href
  {\doibase 10.1016/S0370-2693(00)00669-9} {\bibfield  {journal} {\bibinfo
  {journal} {Phys. Lett. B}\ }\textbf {\bibinfo {volume} {485}},\ \bibinfo
  {pages} {208} (\bibinfo {year} {2000})},\ \Eprint
  {http://arxiv.org/abs/hep-th/0005016} {arXiv:hep-th/0005016} \BibitemShut
  {NoStop}%
\bibitem [{\citenamefont {Hu}\ and\ \citenamefont {Sawicki}(2007)}]{Hu:2007nk}%
  \BibitemOpen
  \bibfield  {author} {\bibinfo {author} {\bibfnamefont {W.}~\bibnamefont
  {Hu}}\ and\ \bibinfo {author} {\bibfnamefont {I.}~\bibnamefont {Sawicki}},\
  }\href {\doibase 10.1103/PhysRevD.76.064004} {\bibfield  {journal} {\bibinfo
  {journal} {Phys. Rev. D}\ }\textbf {\bibinfo {volume} {76}},\ \bibinfo
  {pages} {064004} (\bibinfo {year} {2007})},\ \Eprint
  {http://arxiv.org/abs/0705.1158} {arXiv:0705.1158 [astro-ph]} \BibitemShut
  {NoStop}%
\bibitem [{\citenamefont {Koyama}\ \emph {et~al.}(2009)\citenamefont {Koyama},
  \citenamefont {Taruya},\ and\ \citenamefont {Hiramatsu}}]{Koyama_2009}%
  \BibitemOpen
  \bibfield  {author} {\bibinfo {author} {\bibfnamefont {K.}~\bibnamefont
  {Koyama}}, \bibinfo {author} {\bibfnamefont {A.}~\bibnamefont {Taruya}}, \
  and\ \bibinfo {author} {\bibfnamefont {T.}~\bibnamefont {Hiramatsu}},\ }\href
  {\doibase 10.1103/physrevd.79.123512} {\bibfield  {journal} {\bibinfo
  {journal} {Physical Review D}\ }\textbf {\bibinfo {volume} {79}} (\bibinfo
  {year} {2009}),\ 10.1103/physrevd.79.123512}\BibitemShut {NoStop}%
\bibitem [{\citenamefont {Oyaizu}\ \emph {et~al.}(2008)\citenamefont {Oyaizu},
  \citenamefont {Lima},\ and\ \citenamefont {Hu}}]{Oyaizu:2008tb}%
  \BibitemOpen
  \bibfield  {author} {\bibinfo {author} {\bibfnamefont {H.}~\bibnamefont
  {Oyaizu}}, \bibinfo {author} {\bibfnamefont {M.}~\bibnamefont {Lima}}, \ and\
  \bibinfo {author} {\bibfnamefont {W.}~\bibnamefont {Hu}},\ }\href {\doibase
  10.1103/PhysRevD.78.123524} {\bibfield  {journal} {\bibinfo  {journal} {Phys.
  Rev. D}\ }\textbf {\bibinfo {volume} {78}},\ \bibinfo {pages} {123524}
  (\bibinfo {year} {2008})},\ \Eprint {http://arxiv.org/abs/0807.2462}
  {arXiv:0807.2462 [astro-ph]} \BibitemShut {NoStop}%
\bibitem [{\citenamefont {Schmidt}(2009)}]{Schmidt_2009}%
  \BibitemOpen
  \bibfield  {author} {\bibinfo {author} {\bibfnamefont {F.}~\bibnamefont
  {Schmidt}},\ }\href {\doibase 10.1103/physrevd.80.123003} {\bibfield
  {journal} {\bibinfo  {journal} {Physical Review D}\ }\textbf {\bibinfo
  {volume} {80}} (\bibinfo {year} {2009}),\
  10.1103/physrevd.80.123003}\BibitemShut {NoStop}%
\bibitem [{\citenamefont {Winther}\ \emph {et~al.}(2015)\citenamefont {Winther}
  \emph {et~al.}}]{Winther:2015wla}%
  \BibitemOpen
  \bibfield  {author} {\bibinfo {author} {\bibfnamefont {H.~A.}\ \bibnamefont
  {Winther}} \emph {et~al.},\ }\href {\doibase 10.1093/mnras/stv2253}
  {\bibfield  {journal} {\bibinfo  {journal} {Mon. Not. Roy. Astron. Soc.}\
  }\textbf {\bibinfo {volume} {454}},\ \bibinfo {pages} {4208} (\bibinfo {year}
  {2015})},\ \Eprint {http://arxiv.org/abs/1506.06384} {arXiv:1506.06384
  [astro-ph.CO]} \BibitemShut {NoStop}%
\bibitem [{\citenamefont {Cataneo}\ \emph {et~al.}(2019)\citenamefont
  {Cataneo}, \citenamefont {Lombriser}, \citenamefont {Heymans}, \citenamefont
  {Mead}, \citenamefont {Barreira}, \citenamefont {Bose},\ and\ \citenamefont
  {Li}}]{Cataneo:2018cic}%
  \BibitemOpen
  \bibfield  {author} {\bibinfo {author} {\bibfnamefont {M.}~\bibnamefont
  {Cataneo}}, \bibinfo {author} {\bibfnamefont {L.}~\bibnamefont {Lombriser}},
  \bibinfo {author} {\bibfnamefont {C.}~\bibnamefont {Heymans}}, \bibinfo
  {author} {\bibfnamefont {A.}~\bibnamefont {Mead}}, \bibinfo {author}
  {\bibfnamefont {A.}~\bibnamefont {Barreira}}, \bibinfo {author}
  {\bibfnamefont {S.}~\bibnamefont {Bose}}, \ and\ \bibinfo {author}
  {\bibfnamefont {B.}~\bibnamefont {Li}},\ }\href {\doibase
  10.1093/mnras/stz1836} {\bibfield  {journal} {\bibinfo  {journal} {Mon. Not.
  Roy. Astron. Soc.}\ }\textbf {\bibinfo {volume} {488}},\ \bibinfo {pages}
  {2121} (\bibinfo {year} {2019})},\ \Eprint {http://arxiv.org/abs/1812.05594}
  {arXiv:1812.05594 [astro-ph.CO]} \BibitemShut {NoStop}%
\bibitem [{\citenamefont {{Ntelis}}\ \emph {et~al.}(2017)\citenamefont
  {{Ntelis}} \emph {et~al.}}]{Ntelis:2017nrj}%
  \BibitemOpen
  \bibfield  {author} {\bibinfo {author} {\bibfnamefont {P.}~\bibnamefont
  {{Ntelis}}} \emph {et~al.},\ }\href {\doibase 10.1088/1475-7516/2017/06/019}
  {\bibfield  {journal} {\bibinfo  {journal} {\jcap}\ }\textbf {\bibinfo
  {volume} {2017}},\ \bibinfo {eid} {019} (\bibinfo {year} {2017})},\ \Eprint
  {http://arxiv.org/abs/1702.02159} {arXiv:1702.02159 [astro-ph.CO]}
  \BibitemShut {NoStop}%
\bibitem [{\citenamefont {Sobol'}(2001)}]{SOBOL2001271}%
  \BibitemOpen
  \bibfield  {author} {\bibinfo {author} {\bibfnamefont {I.}~\bibnamefont
  {Sobol'}},\ }\href {\doibase https://doi.org/10.1016/S0378-4754(00)00270-6}
  {\bibfield  {journal} {\bibinfo  {journal} {Mathematics and Computers in
  Simulation}\ }\textbf {\bibinfo {volume} {55}},\ \bibinfo {pages} {271}
  (\bibinfo {year} {2001})},\ \bibinfo {note} {the Second IMACS Seminar on
  Monte Carlo Methods}\BibitemShut {NoStop}%
\bibitem [{\citenamefont {A}\ \emph {et~al.}(2008)\citenamefont {A},
  \citenamefont {M}, \citenamefont {T}, \citenamefont {F}, \citenamefont {J},
  \citenamefont {D}, \citenamefont {M},\ and\ \citenamefont {S}}]{JRC40639}%
  \BibitemOpen
  \bibfield  {author} {\bibinfo {author} {\bibfnamefont {S.}~\bibnamefont {A}},
  \bibinfo {author} {\bibfnamefont {R.}~\bibnamefont {M}}, \bibinfo {author}
  {\bibfnamefont {A.}~\bibnamefont {T}}, \bibinfo {author} {\bibfnamefont
  {C.}~\bibnamefont {F}}, \bibinfo {author} {\bibfnamefont {C.}~\bibnamefont
  {J}}, \bibinfo {author} {\bibfnamefont {G.}~\bibnamefont {D}}, \bibinfo
  {author} {\bibfnamefont {S.}~\bibnamefont {M}}, \ and\ \bibinfo {author}
  {\bibfnamefont {T.}~\bibnamefont {S}},\ }\href
  {http://eu.wiley.com/WileyCDA/WileyTitle/productCd-0470059974.html} {\emph
  {\bibinfo {title} {Global Sensitivity Analysis: The Primer}}}\ (\bibinfo
  {publisher} {Wiley},\ \bibinfo {address} {Chichester (England)},\ \bibinfo
  {year} {2008})\BibitemShut {NoStop}%
\bibitem [{\citenamefont {Sobol'}(1967)}]{SOBOL196786}%
  \BibitemOpen
  \bibfield  {author} {\bibinfo {author} {\bibfnamefont {I.}~\bibnamefont
  {Sobol'}},\ }\href {\doibase https://doi.org/10.1016/0041-5553(67)90144-9}
  {\bibfield  {journal} {\bibinfo  {journal} {USSR Computational Mathematics
  and Mathematical Physics}\ }\textbf {\bibinfo {volume} {7}},\ \bibinfo
  {pages} {86} (\bibinfo {year} {1967})}\BibitemShut {NoStop}%
\bibitem [{\citenamefont {Mckay}\ \emph {et~al.}(2000)\citenamefont {Mckay},
  \citenamefont {Beckman},\ and\ \citenamefont {Conover}}]{Mckay01022000}%
  \BibitemOpen
  \bibfield  {author} {\bibinfo {author} {\bibfnamefont {M.~D.}\ \bibnamefont
  {Mckay}}, \bibinfo {author} {\bibfnamefont {R.~J.}\ \bibnamefont {Beckman}},
  \ and\ \bibinfo {author} {\bibfnamefont {W.~J.}\ \bibnamefont {Conover}},\
  }\href {\doibase 10.1080/00401706.2000.10485979} {\bibfield  {journal}
  {\bibinfo  {journal} {Technometrics}\ }\textbf {\bibinfo {volume} {42}},\
  \bibinfo {pages} {55} (\bibinfo {year} {2000})}\BibitemShut {NoStop}%
\bibitem [{\citenamefont {Ntelis}\ \emph {et~al.}(2018)\citenamefont {Ntelis},
  \citenamefont {Ealet}, \citenamefont {Escoffier}, \citenamefont {Hamilton},
  \citenamefont {Hawken}, \citenamefont {Le~Goff}, \citenamefont {Rich},\ and\
  \citenamefont {Tilquin}}]{Ntelis:2018ctq}%
  \BibitemOpen
  \bibfield  {author} {\bibinfo {author} {\bibfnamefont {P.}~\bibnamefont
  {Ntelis}}, \bibinfo {author} {\bibfnamefont {A.}~\bibnamefont {Ealet}},
  \bibinfo {author} {\bibfnamefont {S.}~\bibnamefont {Escoffier}}, \bibinfo
  {author} {\bibfnamefont {J.-C.}\ \bibnamefont {Hamilton}}, \bibinfo {author}
  {\bibfnamefont {A.~J.}\ \bibnamefont {Hawken}}, \bibinfo {author}
  {\bibfnamefont {J.-M.}\ \bibnamefont {Le~Goff}}, \bibinfo {author}
  {\bibfnamefont {J.}~\bibnamefont {Rich}}, \ and\ \bibinfo {author}
  {\bibfnamefont {A.}~\bibnamefont {Tilquin}},\ }\href {\doibase
  10.1088/1475-7516/2018/12/014} {\bibfield  {journal} {\bibinfo  {journal}
  {JCAP}\ }\textbf {\bibinfo {volume} {12}},\ \bibinfo {pages} {014} (\bibinfo
  {year} {2018})},\ \Eprint {http://arxiv.org/abs/1810.09362} {arXiv:1810.09362
  [astro-ph.CO]} \BibitemShut {NoStop}%
\end{thebibliography}%

\end{document}